\documentclass[jkps,preprint,fleqn,showpacs,showkeys]{revtex4-1}
\usepackage{graphicx}
\usepackage{amssymb}
\usepackage{amsmath}
\usepackage{bm}
\usepackage{gt11}
\usepackage{color}
\begin{document}
\newcommand{\jvm}{\jv_{\rm m}}
\newcommand{\rhom}{\rho_{\rm m}}
\newcommand{\jH}{j_{\rm h}}
\newcommand{\jHv}{\jv_{\rm h}}
\newcommand{\jvH}{\jv_{\rm h}}
\newcommand{\Qv}{{\bm Q}}
\newcommand{\rhoH}{\rho_{\rm h}}
\newcommand{\rhoh}{\rho_{\rm h}}
\newcommand{\sigmac}{\sigma_{\rm c}}
\setcounter{page}{0}
\title[]{
Monopoles in ferromagnetic metals 
}
\author{Gen \surname{Tatara}}
\email{tatara@tmu.ac.jp}
\author{Akihito \surname{Takeuchi}}
\author{Noriyuki \surname{Nakabayashi}}\author{Katsuhisa  \surname{Taguchi}}
\affiliation{Graduate School of Science, 
Tokyo Metropolitan University, 1-1 Minamiosawa, Hachioji
192-0397, Japan}

\date[]{Received }

\begin{abstract}
The aim of this short review is to give an introduction to monopoles and to present theoretical derivation of two particular monopoles in ferromagnetic metals, a hedgehog monopole and a spin damping monopole.
In an electromagnetism in the vacuum, described by Maxwell's equations, magnetic field and electric field are not symmetric, since there is no monopole, a particle having a finite magnetic charge.
Still, monopole has been an exciting object for a long time, and was discussed on a phenomenological ground by Dirac in 1931.
A theoretical possibility of monopole generation was first given by 't Hooft and Polyakov in 1974 in a context of a symmetry breaking in a grand unified theory (GUT).
A GUT monopole has not been discovered in experiments so far.
In contrast to in the vacuum, several kinds of monopoles are expected to emerge in solids associated with various symmetry breaking mechanisms.
Of particular interest is metallic ferromagnetic systems, since a breaking of a symmetry of conduction electron spin, described by an SU(2) algebra, can give rise to monopoles.
Indeed two monopoles are theoretically predicted in ferromagnets; one is a hedgehog monopole arising from a topological spin structure and the other arising from spin damping in the presence of spin-orbit interaction.
In this paper, we focus on these monopoles, while other objects similar to monopoles but not coupled to effective electromagnetic fields, such as spin ice monopoles, are touched only briefly in the introduction. 
Those monopoles are extended objects coupled to effective electromagnetic fields, which are described by Maxwell's equations with monopole contribution.
The effective fields are the one coupled to spin of a particle like electron, muon and neutron; 
two monopoles are thus detectable by electric measurements.
Spin damping monopoles can be generated in simple systems such as a junction of a ferromagnet and a heavy element with strong spin-orbit interaction such as Pt.
This monopole is essential in coupling electronics with magnetism, and is thus expected to play an essential role in spintronics.

\end{abstract}



\maketitle

\section{Introduction}
\subsection{Monopole}

\begin{center}
\begin{figure}
\includegraphics[width=0.3\hsize]{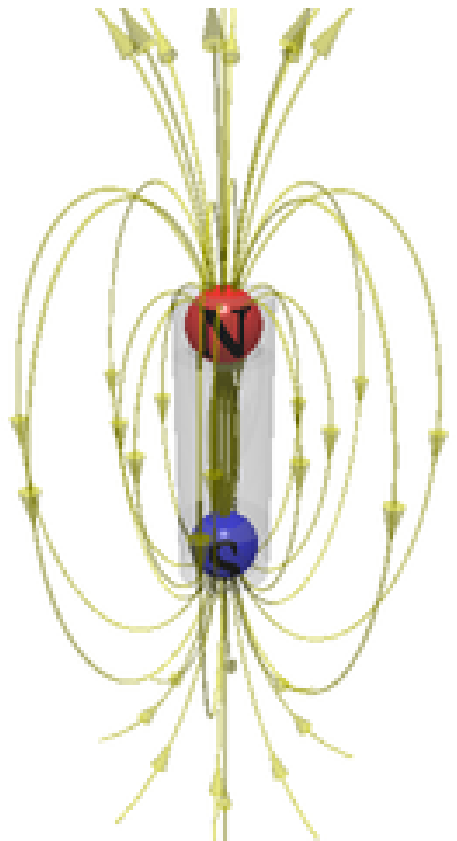}
\includegraphics[width=0.3\hsize]{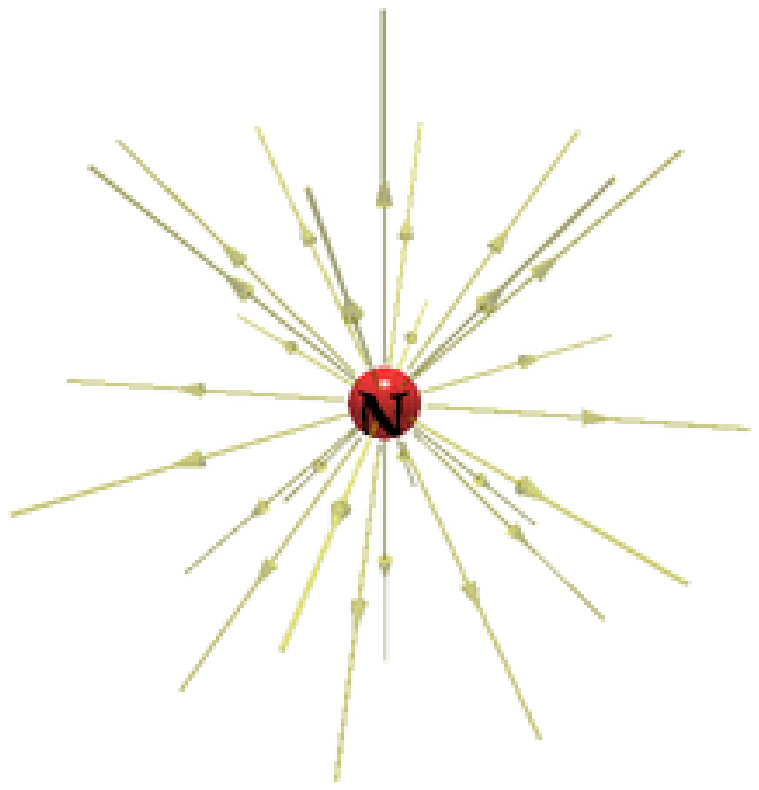}
\caption{
Left: Schematic figure of a magnet and magnetic field around it. 
Right: A monopole, particle having only N (or S) magnetic pole, and the magnetic field it emits.
\label{fig:dipole} }

\end{figure}
\end{center}

Macroscopic magnets have two poles called N (north) and S (south) at two different edges (Fig. \ref{fig:dipole}).
We also know that we cannot extract only N or S pole from magnets;
when we divide a magnet, a pair of N and S poles is created at the edge and each piece becomes a magnet with the equal amount of N and S poles.
There is therefore no monopole at least in the energy scale we concern in our life.
This is because the magnets are made of spin, a quantum magnet the electron carries, which creates the divergencelss  magnetic field only.
In other words, each spin contain N and S poles, which are not separable since elementary particles are pointlike.

Possibility of a particle carrying only N or S pole, called magnetic monopole, was first discussed by Dirac \cite{Dirac31}. 
He showed based on a macroscopic consideration that monopole is tied to a string of singularity, and its existence is allowed only if the charge of the monopole is quantized by unit of $2\pi\hbar/e$ ($e$ is the electron charge).
A microscopic mechanism of monopole creation was theoretically discovered by 
't Hooft \cite{tHooft74} and Polyakov \cite{Polyakov74} independently.
They showed that monopole arises in a grand unified theory (GUT) of elementally particles when the symmetry breaking of the SU(5) symmetry to the U(1) symmetry of the electromagnetic field occurs. 
Such monopoles may have been created in the early universe (at about $10^{-10}$ sec after the big bang) and may still be around us.
So far, however, no evidence has been obtained in experiments waiting for a monopole from universe
to go through superconducting detectors \cite{Cabrera82} or detecting the ionization \cite{macro02}.
The energy scale of the GUT monopole is $10^{17}$ GeV, and so creating one in accelerator on the earth is impossible.

\subsection{Maxwell's equations without monopole}

Before starting to discuss monopoles, let us discuss the law of conventional electromagnetism without monopole.
We know that electric field, $\Ev$, and magnetic field, $\Bv$, behave differently, i.e., electric field has finite divergence while magnetic field is divengenceless \cite{Jackson98}.
To put in equations,
\begin{align}
\nabla\cdot\Ev&=\frac{\rho}{\epsilon} \nnr
\nabla\cdot\Bv&=0,  \label{divergence}
\end{align} 
where $\rho$ is charge density and $\epsilon$ is dielectric constant.
The two fields are governed by another set of equations describing rotational components;
\begin{align}
\nabla\times\Ev&=-\delp{\Bv}{t} \nnr
\nabla\times\Bv&= \mu\jv+\epsilon \mu \delp{\Ev}{t}, \label{rotation}
\end{align} 
where $\mu$ is magnetic permeability and $\jv$ is current density.
The fields $\Ev$ and $\Bv$ are therefore not symmetric; magnetic field is generated by electric current by the Amp\`ere's law, while there is no driving current for electric field.
One may imagine that the electromagnetism law is more beautiful  
if there is a current that drives electric field via an analog of the Amp\`ere's law.
Let us write such current as $\jvm$. The first equation of Eq. (\ref{rotation}) then becomes
\begin{align}
\nabla\times\Ev&=-\jvm-\delp{\Bv}{t} .\label{rotE}
\end{align} 
Taking the divergence of this equation, we have
\begin{align}
0&=-\nabla\cdot\jvm-\delpo{t}(\nabla\cdot\Bv) ,
\end{align} 
which indicates that the second equation of Eq. (\ref{divergence}) needs to be modified as
\begin{align}
\nabla\cdot\Bv&=\rhom , \label{divB}
\end{align} 
where $\rhom$ is a quantity  satisfying a conservation law of
\begin{align}
\dot{\rhom}+\nabla\cdot\jvm=0. \label{mpconservation}
\end{align} 
Equation (\ref{divB}) indicates that $\rhom$ emits a magnetic field like a N or S pole; 
hence $\rhom$ is a density of monopole.
The current $\jvm$ is then a monopole current.
Thus, if monopole exists, $\Ev$ and $\Bv$ become symmetric, and the electric field can be generated by applying the monopole current.
\begin{center}
\begin{figure}
\includegraphics[width=0.3\hsize]{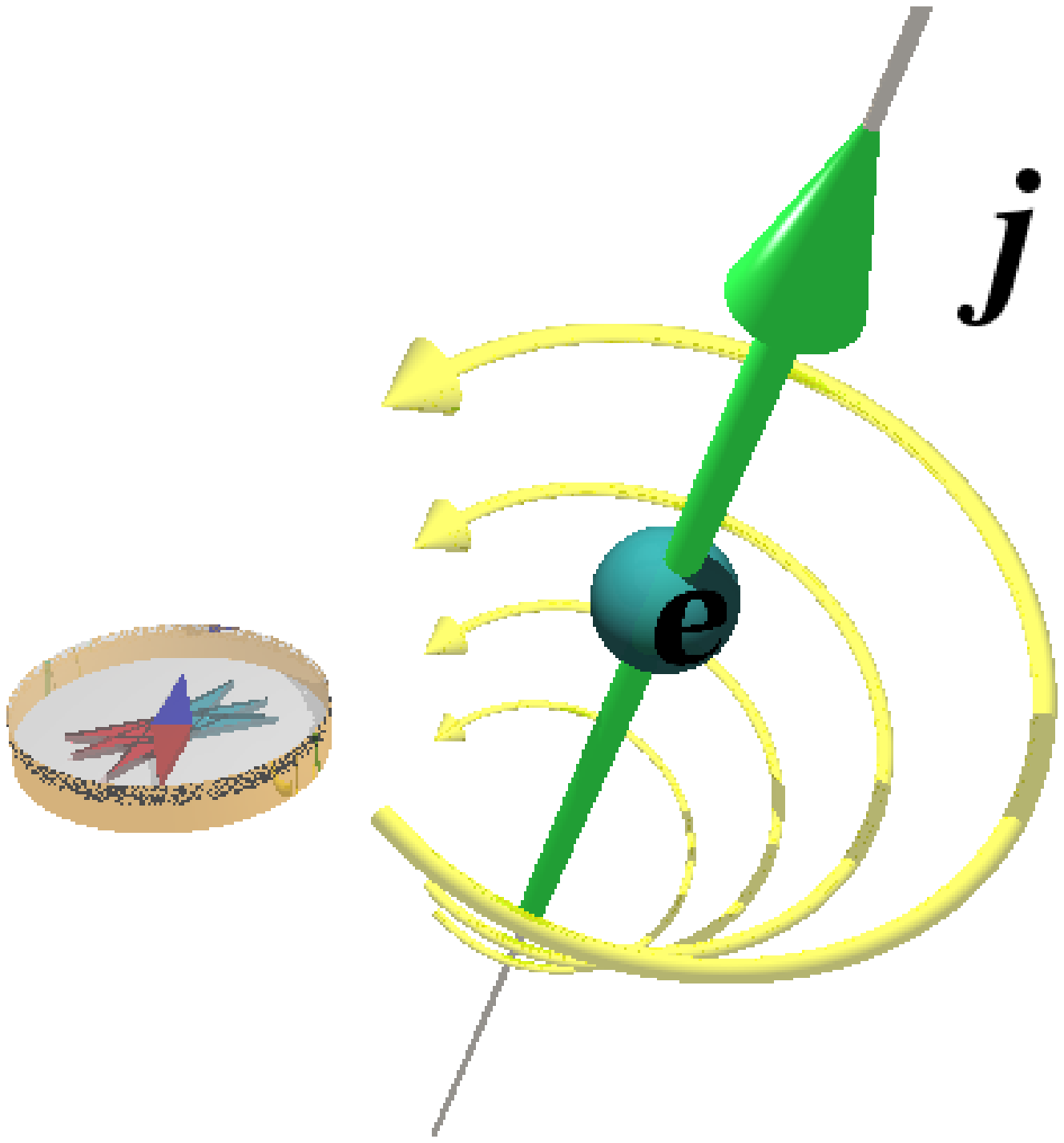}
\includegraphics[width=0.3\hsize]{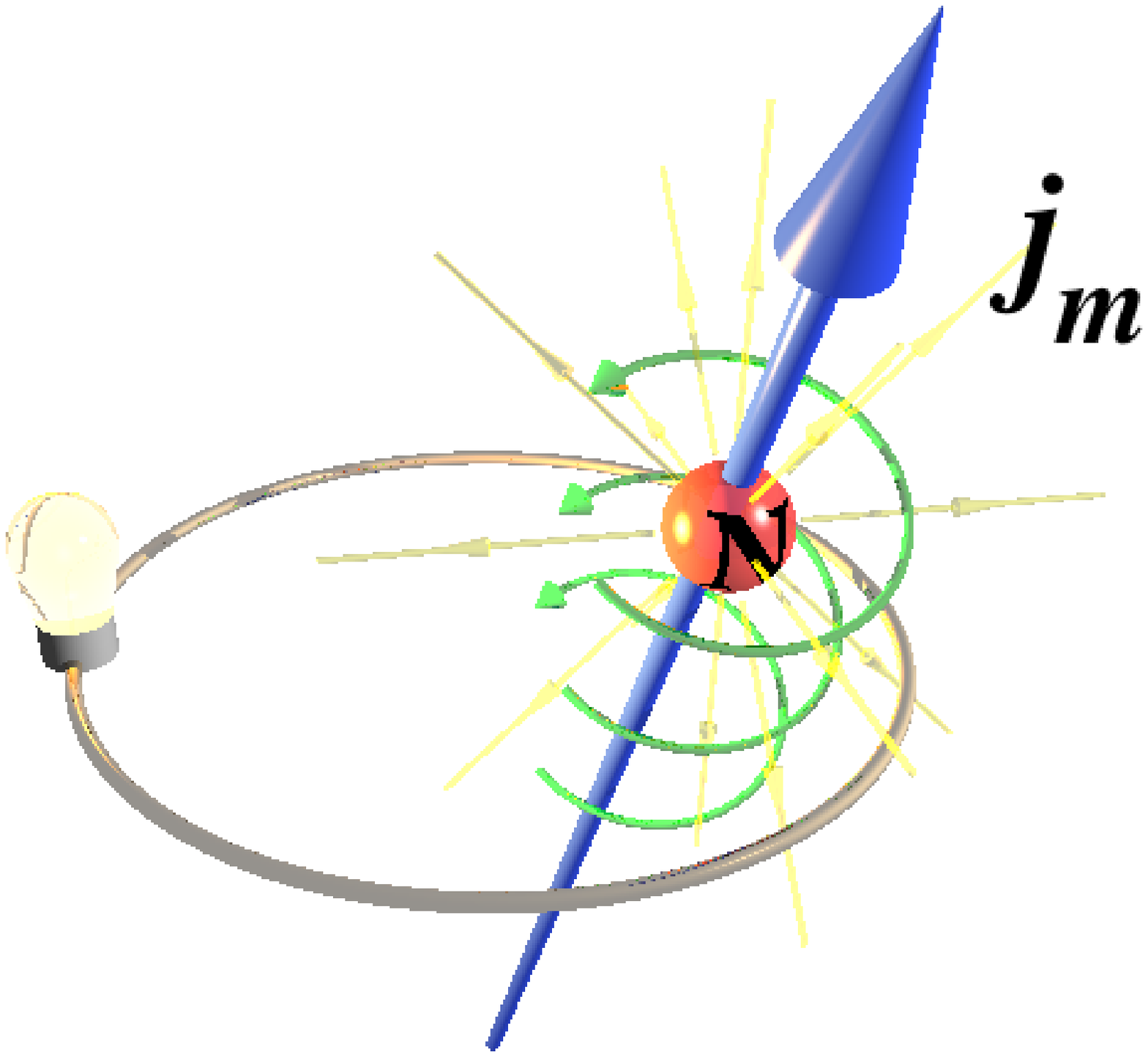}
\caption{
Left) The Amp\`ere's law allows us to generate magnetic field  from the electric current, $\jv$. 
In contrast, there is no Amp\`ere's law for the electric field in the electromagnetism if U(1) symmetry is exact.
The exception is the case with monopole: monopole current, $\jvm$, generates $\Ev$ (Right). }
\label{fig:ampere}
\end{figure}
\end{center}

The aim of this short review is to give an introductory description of monopoles in condensed matter and explain in detail monopoles in ferromagnetic metals.
The paper is organized as follows. 
Monopole is discussed from a macroscopic viewpoint in \S\ref{sec:dirac}.
Section \ref{sec:mpinsolid} is a brief introduction of monopoles in solids.
Other topological objects similar to monopoles are introduced in \S\ref{sec:topologicalobjects}.
These objects are touched only briefly, since we would like in this paper to focus on monopoles which really couple to electromagnetism via Maxwell's equations.  
Sections \ref{sec:hhmp} and \ref{sec:sdmp} are the main content of the paper presenting theoretical description of hedgehog monopole and spin damping monopole, respectively.
Relativistic notation is summarized in \S\ref{sec:relativistic} and spin damping is briefly described in \S\ref{sec:damping}.

\section{Monopole from macroscopic viewpoint : Dirac's string
\label{sec:dirac}}
Let us go on to discuss monopoles from a macroscopic viewpoint, i.e., based on the Maxwell's equations. 
As is well-known, in the electromagnetism without monopole, the magnetic field $\Bv$ is expressed as a rotation of a vector potential, namely, as $\Bv=\nabla \times \Av$.
When monopole is present, this is no longer true, but still one can define a vector potential as a line integral over a certain line $L$ starting from monopole to the infinity.
Let us consider a case of a single monopole at $\rv=0$, 
\begin{align}
 \Bv(\rv)=\frac{g}{4\pi}\frac{\rv}{r^3},
\end{align}
where $g$ is a monopole charge.
We can define a "vector potential" 
\begin{align}
A_i^{(L)}(\rv)=\sum_{jk}\epsilon_{ijk}\int_Ldr'_j B_k(\rv-\rv'), \label{AfromBstring}
\end{align}
where $B_k$ is the $k$th component of the magnetic field (including monopole) and $L$ is a string connecting $\rv=0$ and the infinity.
The "magnetic field" of this vector potential $\Av^{(L)}$ then reads
\begin{align}
\nabla\times \Av^{(L)}(\rv)=\Bv(\rv)+\int_L d\rv' \rhom(\rv-\rv'),
\label{nablaAL}
\end{align}
where $\rhom$ is the monopole density,  $\rhom(\rv)=g\delta^3(\rv)$ if a single monopole case.
The first term of Eq. (\ref{nablaAL}) correctly reproduces the field $\Bv$ but we have an additional non local singular field represented by the second term.
The singular magnetic field represented by the second term of \Eqref{nablaAL} exists along the string $L$.
Dirac argued therefore that monopole needs to have such string (called the Dirac's string) but this string must not been "seen" by physical particles \cite{Dirac31}. 
If many monopoles exist, many strings $L$ each attached to a single monopole are needed to describe the vector potential.

\begin{figure}[tb]
\begin{center}
\includegraphics[width=0.3\hsize]{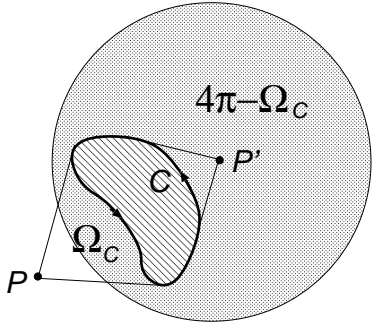}
\end{center}
\caption{
When a solid angle of a closed path $C$ observed at P is $\Omega_C$, it is $4\pi-\Omega_C$ at P' in the opposite side of the path. 
\label{fig:solidangle}
}

\end{figure}

The unobservability of the string is ensured if the monopole charge $g$ is quantized to certain values.
One way to derive the quantization condition is to impose the condition that the shift of the Dirac's string does not modify the wave function of the charged particles \cite{Jackson98}. 
Equation (\ref{AfromBstring}), explicitely written in a single monopole case reads
\begin{align}
\displaystyle A_i^{(L)}(\rv)=\frac{g}{4\pi}\sum_{jk}\epsilon_{ijk}\int_L dr'_j\frac{(r-r')_k}{|\rv-\rv'|^3}.
\end{align}
If one shift the string $L$ to be another string $L'$, the vector potential changes to
\begin{align}
 A_i^{(L')}(\rv)=A_i^{(L)}(\rv)+\frac{g}{4\pi}\sum_{jk}\epsilon_{ijk}\int_C dr'_j\frac{(r-r')_k}{|\rv-\rv'|^3},
\label{ALALp}
\end{align}
where $C=L'-L$ is a closed path surrounded by $L$ and $L'$.
The last term of the right-hand side is proportional to the derivative of the solid angle $\Omega_C(\rv)$ subtended by $C$ observed at $\rv$.
In fact, a difference of solid angles observed at $\rv$ and $\rv+\delta\rv$, which we call $\delta \Omega_C$, 
is (noting that $d\rv' \times \delta\rv$ is a vector normal to the plane having a length of an element of area)
\begin{align}
\delta \Omega_C=
 \int_C (d\rv' \times \delta\rv)\cdot \frac{\rv'-\rv}{|\rv-\rv'|^3}.
\end{align}
We therefore see that 
\begin{align}
\nabla \Omega_C=
 \int_C d\rv' \times \frac{(\rv-\rv')}{|\rv-\rv'|^3},
\end{align}
and that \Eqref{ALALp} becomes 
\begin{align}
 \Av^{(L')}=\Av^{(L)}+\frac{g}{4\pi}\nabla \Omega_C.
\end{align}
Since a phase for the electron's wave function is given by an integral of the vector potential, 
this modification of the vector potential results in a phase change of 
$e^{\frac{ie}{\hbar}\frac{g}{4\pi}\Omega_C}$.
Quantum mechanics requires that this phase is single-valued.
However, solid angle $\Omega_C$ is not single-valued. 
Multi-valuedness of $\Omega_C$  occurs when $\Omega_C$ changes by $4\pi$ as an electron goes through the plane spanned by $C$ (Fig. \ref{fig:solidangle}).
The single-valuedness of the electron's phase is thus ensured by requiring that 
\begin{align}
{eg}=2\pi n \hbar , \label{diracquantization}
\end{align}
 where $n$ is an integer.
This is the quantization condition for the monopole charge, pointed out by Dirac.

The above argument is based on the assumption that the magnetic field is written as a rotation of a U(1) vector potential even in the presence of monopoles.
This assumption, however, is wrong, since the exact U(1) gauge invariance does not exist when monopole exists. 
As we will show later, the monopole field in the case of the symmetry breaking mechanism is expressed by the additional field orthogonal to the U(1) degrees of freedom. 
The Dirac's string containing a singular magnetic field (the last term of \Eqref{nablaAL}), is thus an artifact arising when one tries to describe the monopole field within the U(1) gauge theory assuming $\Bv=\nabla\times \Av$, and it is eliminated when one takes account of microscopic origin of the monopole.

In the case of 't Hooft-Polyakov monopole, monopole is created from the symmetry breaking \cite{Ryder96}.
Their argument is for a system of non-Abelian gauge field coupled to a Higgs represented by a vector field, $\phi^\alpha$ ($\alpha=x,y,z$). 
They showed that there is a solution where Higgs field behaves at the infinity as
\begin{align}
\phi^\alpha \propto \frac{r^\alpha}{r}  \;\;\; (r\ra \infty).
\end{align}
In this configuration, magnetic field is shown to contain monopole, and monopole density is determined by Higgs field as
\begin{align}
\rhom\propto \sum_{ijk}\sum_{\alpha\beta\gamma} \epsilon_{ijk}\epsilon_{\alpha\beta\gamma}
\partial_i \phi^\alpha \partial_j \phi^\beta \partial_k \phi^\gamma.
\end{align}
A volume integral of this monopole density is written as a surface integral at the infinity, and it turns out to take only integer value called a winding number.
This topological nature of the Higgs field leads to a quantization of monopole charge. 
Thus, 't Hooft-Polyakov monopole emerges from a symmetry breaking from non-Abelian group to U(1) group driven by a condensation of Higgs field.
Dirac's string is not necessary here, since monopole field is created from Higgs field, which is different from a gauge field.

In \S\ref{sec:hhmp}, we will discuss in detail a monopole arising from a symmetry breaking in ferromagnets.

\section{Monopoles in solids
\label{sec:mpinsolid}}

\subsection{Hedgehog monopole}

As we have mentioned, in the electromagnetism in the vacuum, monopole predicted at high energy has not been found.  
Even if it is found in the future, we cannot make a device since its energy is too high.
In contrast to electromagnetism in the vacuum, we have hope in solids.
In fact, electrons in solids feel another U(1) gauge field which couples to the electron's spin \cite{Stern92}.
The spins are object in an SU(2) space, and thus if the breaking of its symmetry occurs, a resulting effective U(1) gauge field may contain monopoles according to the 't Hooft and Polyakov scenario.
In addition, the energy scale of the symmetry breaking is at low energy, less than 1eV, and so device application would be straightforward.

\begin{figure}
\begin{center}
\includegraphics[scale=0.35]{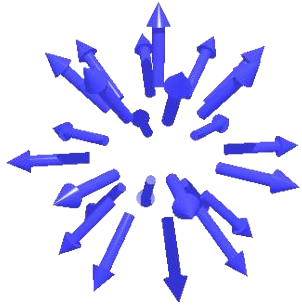}
\end{center}
\caption{
Schematic illustration of a magnetization structure for a hedgehog monopole, discussed by Volovik.
This structure has a singularity at the center, and is topologically non-trivial. 
The electron coupled to this magnetization structure feels the effective electromagnetic field with a monopole.
}
\label{fig:hedgehog}
\end{figure}

Such a monopole was indeed theoretically pointed out to exist in ferromagnetic metals by Volovik in 1987 \cite{Volovik87}.
In fact, a monopole arises from a strong sd coupling between the conduction electron and local spin, which specifies the projection of the conduction electron spin having SU(2) symmetry to a U(1) space.
When local spin structure is a topologically non-trivial one called a hedgehog, the projected effective electromagnetic field contains monopoles.

\subsection{Spin damping monopole}

Recently, another monopole in ferromagnetic metals was predicted \cite{Takeuchi12,Takeuchi_MMM12}.
The idea is to include spin-orbit interaction besides sd interaction.
Spin-orbit interaction modifies the projection to the U(1) plane defined by the sd interaction, and thus 
new monopole may arise when the spin-orbit interaction is included.
The monopole generation in such case cannot be discussed by use of gauge fields,
since the spin-orbit interaction is not a gauge interaction.
Instead, novel method based on a transport calculation was applied by Takeuchi and Tatara \cite{Takeuchi12}.
They derived effective electromagnetic fields by calculating the electric current by use of Keldysh Green's functions, and showed that the fields satisfy Maxwell's equation with monopole.
In their unique approach, a knowledge of gauge invariance was not necessary to explore the structure of the electromagnetism.
This fact may sound surprising, but is natural, since a U(1) gauge invariance is equivalent to charge conservation law, which is strictly observed in transport calculations. 
In their case, a deviation from exact U(1) invariance due to spin-orbit interaction resulted in a monopole contribution.
Based on the transport method, they revealed that monopole arises when the spin structure is dynamic and when spin-orbit interaction is included.
More specifically, monopole arises when there is a damping of spin, represented by a damping vector (see \S\ref{sec:damping}), 
\begin{align}
\Nv\equiv \Sv\times \dot{\Sv},
\end{align}
 ($\Sv$ being local spin), and thus the monopole was named the spin damping monopole.

\begin{figure}[tb]
\begin{center}
\includegraphics[width=0.3\hsize]{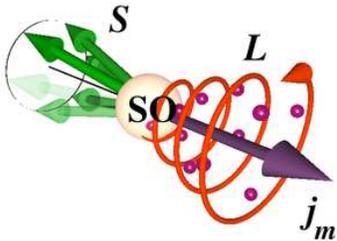}
\end{center}
\caption{
Schematic illustration of a spin damping monopole. 
A damping of spin ($\Sv$) generates an orbital angular momentum $\Lv$ for electrons when spin-orbit interaction (SO) acts. 
A circular motion due to the orbital angular momentum is equivalent to that caused by an electric field induced by a monopole current $\jvm$ via Amp\`ere's law.
}
\label{fig:sdmp}
\end{figure}
A physical mechanism of spin damping monopole is understood as a transfer of the spin angular momentum to  orbital one (Fig. \ref{fig:sdmp}).
In fact, a vector $\alpha \Nv$, where $\alpha$ is the Gilbert damping parameter 
(proportional to the strength of the spin-orbit interaction if in metals), represents the spin angular momentum dissipated. 
This lost angular momentum is converted into the orbital motion of the electrons by the spin-orbit interaction, inducing the circular orbital motion of the electrons.
This circular motion is regarded as a result of a fictitious magnetic field due to a monopole.
>From this intuitive explanation, the monopole density should appear when the spin damping is spatially inhomogeneous.
This is indeed consistent with the result of Ref. \cite{Takeuchi12}, which showed that the monopole density is $\rhom\propto \nabla\cdot\Nv$.
Similarly, when the spin damping is time-dependent, a monopole current arises since a temporal change of  effective magnetic field is equivalent to a monopole current. 
The spin damping monopole was argued to be essential in the spin-charge conversion in spintronics phenomena such as the inverse spin Hall effect \cite{Saitoh06}.

These monopoles in ferromagnetic systems are discussed in detail in \S\ref{sec:hhmp} and \S\ref{sec:sdmp}.

\begin{figure}[tb]
\begin{center}
\includegraphics[width=0.3\hsize]{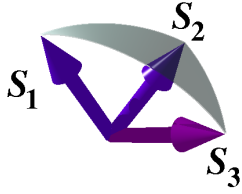}
\end{center}
\caption{
A scalar spin chirality subtended by three spins represents non-coplanarity.
Spin chirality reduces to spin Berry's phase in the slowly varying limit.
}
\label{fig:chirality}
\end{figure}

\section{Related topological objects in solids
\label{sec:topologicalobjects}}
\subsection{Spin chirality}
There are several objects closely related to monopoles.
Spin chirality is one example.
In a spin system on a lattice, a non-coplanarity of three spins, $\Sv_1$, $\Sv_2$ and $\Sv_3$, is represented by a scalar product called a scalar chirality, $\chi\equiv \Sv_1\cdot(\Sv_2\times\Sv_3)$ (Fig. \ref{fig:chirality}).
When the spin structure is slowly varying in space, spin chirality 
can be expanded as  
\begin{equation}
\Sv(\xv) \cdot (\Sv(\xv_1)\times\Sv(\xv_2))
\simeq \sum_{\mu\nu}(x_1-x)_\mu(x_2-x)_\nu
\Sv(\xv) \cdot (\partial_\mu \Sv(\xv)\times\partial_\nu \Sv(\xv)),
\end{equation}
and the spin chirality reduces to the spin Berry phase.
In the continuum limit, therefore, a scalar chirality is represend by a vector
\begin{align}
\chi_i\equiv \sum_{jk}\epsilon_{ijk} \Sv\cdot(\nabla_j\Sv\times \nabla_k\Sv), 
\end{align}
where the direction $i$ is orthogonal to a plane the  three spins lie.
This expression of the scalar chirality is proportional to spin Berry's phase or a local effective magnetic field generated by a hedgehog monopole (see \Eqref{EandBHH}). 
One should note, however, that finite local spin chirality does not necessarily mean that a monopole exists; 
monopole exisitence requires a surface integral of the chirality, 
$\int d\Sv\cdot{\bm \chi}$ ($d\Sv$ is an element of surface integral), to be finite, and this is realized only for a three-dimensional hedgehog structure shown in Fig. \ref{fig:hedgehog}.

Even when monopoles do not exit, a finite spin chirality leads to interesting physics, such as inducing chirality-driven exotic anomalous Hall effect \cite{Ye99,TK02}.
When spin structure such as domain wall is dynamic, the effective electric field acting on the electron's spin, the spin motive force, is induced (see \Eqref{EandBHH}).
Spin motive force was detected in the case of moving domain wall \cite{Yang09}.

A monopole was also found in systems of conventional anomalous Hall effect driven by spin-orbit interaction \cite{Nagaosa08,Nagaosa10}.
This monopole is a singularity in the momentum space and is not coupled to the electromagnetism by  Maxwell's equations.

\begin{figure}
\begin{center}
\includegraphics[width=0.3\hsize]{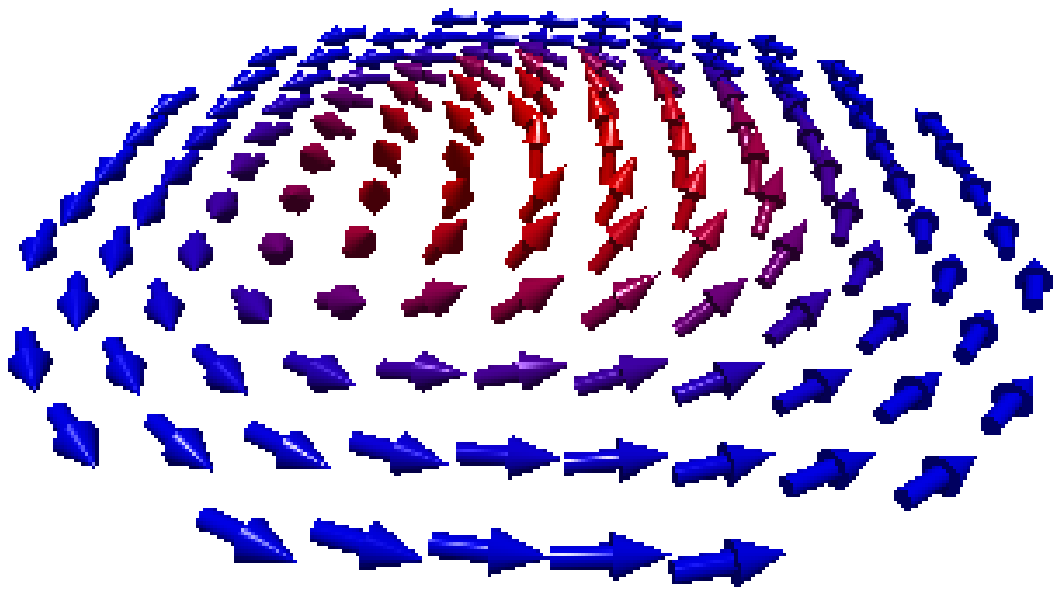}
\includegraphics[width=0.3\hsize]{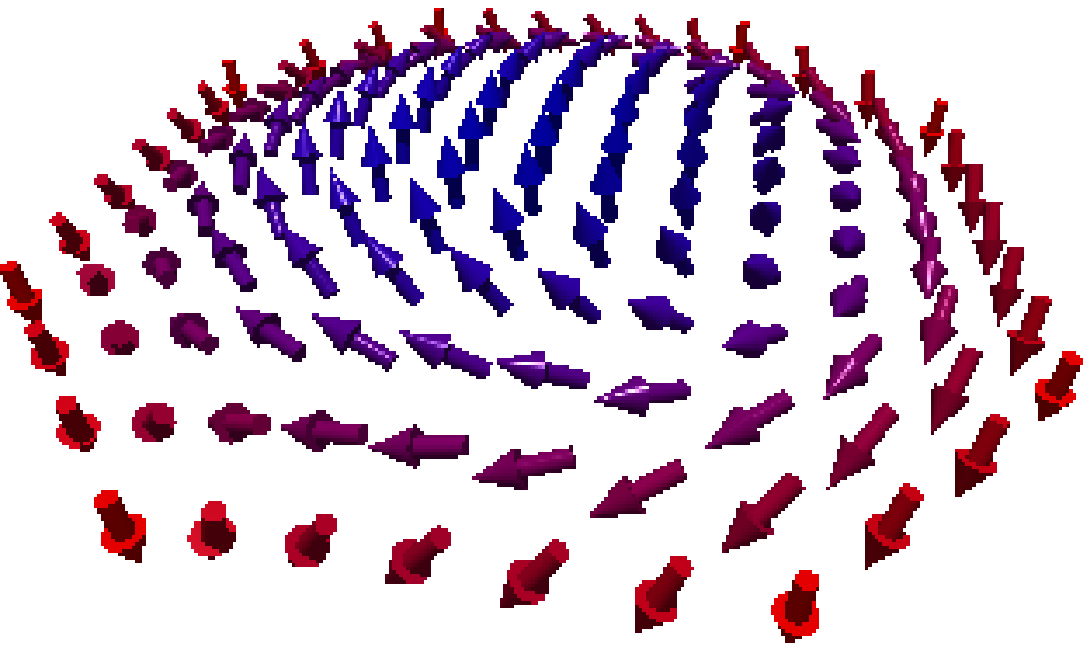}
\end{center}
\caption{
Schematic illustration of a magnetic vortex (left) and a skyrmion (right).
These are two-dimensional structures of magnetization with topological meaning.
The topological number of a vortex is the number of spin rotation as one travels along the perimeter, called a vorticity.
In a skyrmion structure, the spins at the perimeter is fixed to be parallel (spin down in this figure), and its topological number is  the number of spin rotation as one travels from the origin to the infinity along the radial direction.
Vortex and skyrmion in this figure have a topological number of one.
}
\label{fig:vortexandskyrmion}
\end{figure}

\subsection{Vortex and skyrmion \label{sec:vortex}}

Magnetic vortex and skyrmion in magnets are  spin structures having topological feature in two-dimensions (Fig. \ref{fig:vortexandskyrmion}) \cite{Thiele73}.
They have finite value of local spin chirality.
A vortex is a structure where the number of spin rotation as one travels along the circle at the infinity (or the edge of the disk) is the topological number called a vorticity, which is an integer.
Vortex state is stable in a small circular disk of submicron size since the magnetostatic energy at the edge is the lowest for the vortex state \cite{Shinjo00}.
Switching of a vortex core by an electric current is a hot issue from a viewpoint of fundamental science and application to non-volatile memories.
At present, rather large current density of $3.5\times10^{11}$A/m$^2$ is necessary for a core flip, and the switching time is not very fast, about 20ns \cite{Yamada07}.

Skyrmion was originally proposed in high energy physics.
In condensed matter, a magnetization structure  
shown in Fig. \ref{fig:vortexandskyrmion} is called a skyrmion \cite{Rossler06}.
Such a structure was predicted to arise when the inversion symmetry is broken.
Skyrmions forming a lattice  were observed in MnSi \cite{Muhlbauer09,Yu10}.
Vortices and skyrmions have finite effective magnetic field, $\Bv_{\rm s}$, but these topological structures are nothing to do with monopole ($\rhom$ and $\jvm$ vanish), since they are two-dimensional objects.

\begin{figure}
\begin{center}
\includegraphics[width=0.3\hsize]{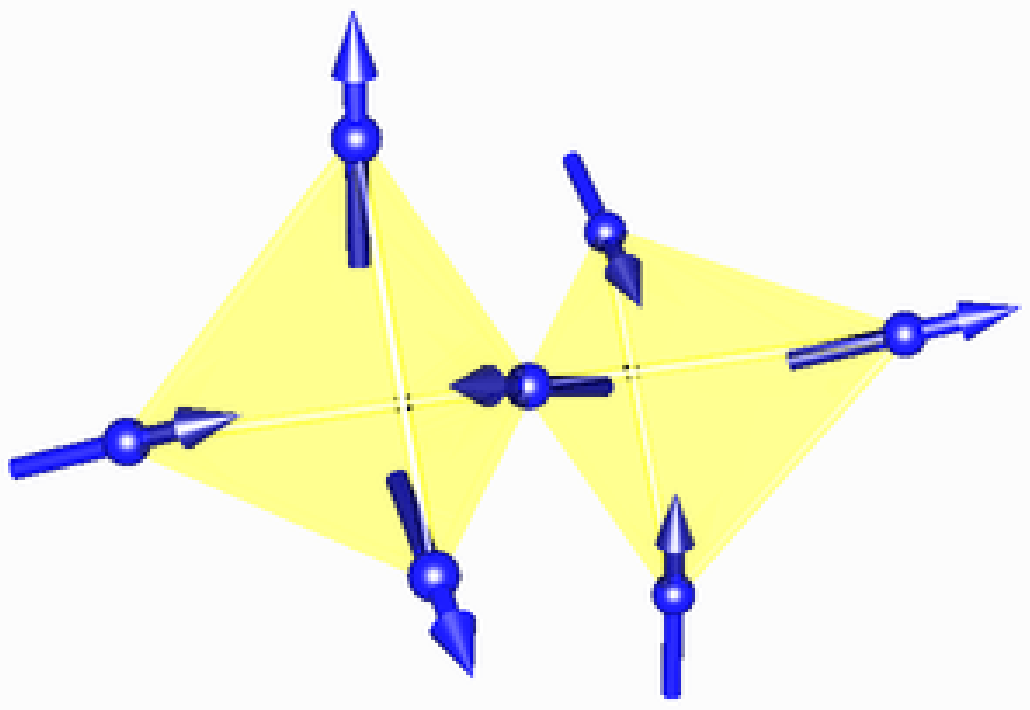}
\includegraphics[width=0.3\hsize]{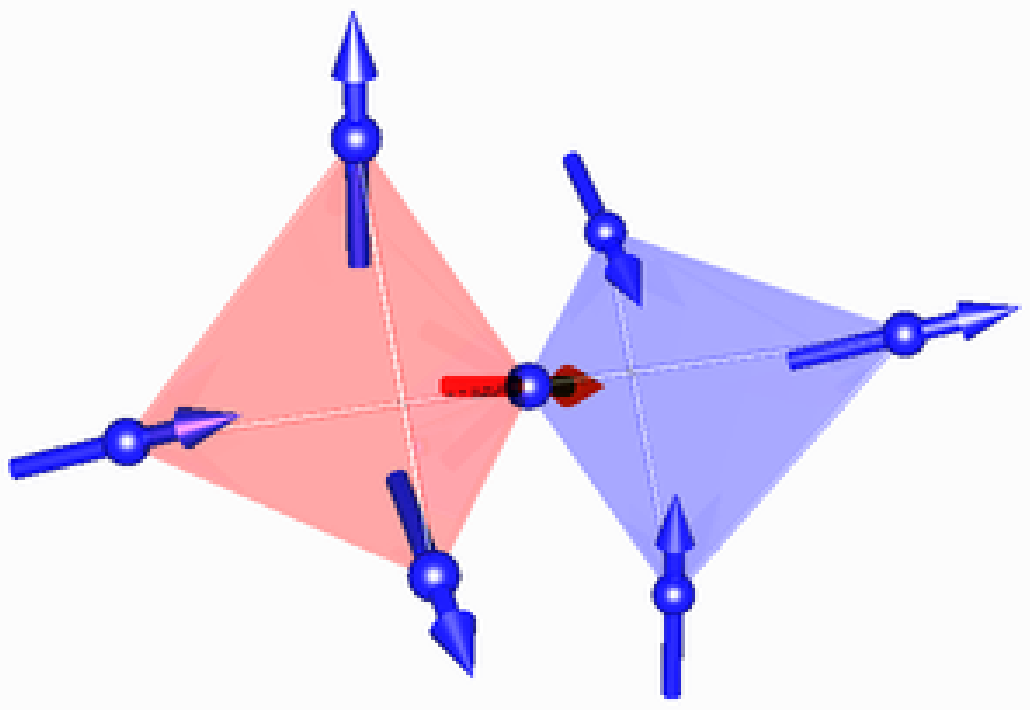}
\end{center}
\caption{
Left: Ground state spin configuration in a frustrated magnet with a pyrochlore lattice. 
All the tetrahedra contains two in-spins and two out-spins, satisfying ice rule.
Right: An excited state of a spin ice. Due to a flip of a spin connecting the two tetrahedra, left tetrahedra has three out-spins and the right tetrahedra has only one out-spin.
This configuration contains effective monopoles; positive and negative ones on the left and right tetrahedra, respectively. 
\label{fig:spinicemonopole}
}
\end{figure}

\subsection{Spin ice monopole}
Monopoles in frustrated magnets with pyrochlore lattice is a hot subject recently \cite{Balents10}.
In the ground state of pyrochlore spin system, each tetrahedra has two spins pointing to the center of tetrahedra (in-spin) and two spins pointing outwards (out-spin), satisfying so called an ice rule. 
When the system is excited, ice rule is broken and some of tetrahedra contain three in-spins and there arise the same number of tetrahedra with three out-spins (Fig. \ref{fig:spinicemonopole}).
In such excited states, tetrahedra with broken ice rule have nonvanishing divergence of magnetization, $\nabla\cdot\Mv\equiv \rho_{M}$.
In terms of magnetic field $\Hv\equiv \frac{1}{\mu_0}\Bv-\Mv$, the excited states of spin ice systems thus have monopoles, i.e., 
 $\nabla\cdot\Hv=- \rho_{M}$ \cite{Ryzhkin05}. 
The spin configuration corresponding to monopoles has been observed by neutron scattering experiment \cite{Kadowaki09}.
In addition, an effective magnetic charge of a spin ice monopole was measured in Dy$_2$Ti$_2$O$_7$ \cite{Bramwell09,Ryzhkin11}.
Spin ice system is also created recently on an artificial square lattice \cite{Morgan11}.

One should understand that spin ice monopole is not a real monopole, since $\nabla\cdot\Bv=0$ is strictly true even in the presence of any spin configurations; it is a monopole of an artificial magnetic field \cite{Balents10}. 
Nevertheless, an spin ice monopole interpretation is highly useful to describe excited states in frustrated spin systems from the viewpoint of spin liquid.

\subsection{Monopole in topological insulator}
Recent study revealed that monopole arises at the surface of topological insulators as a result of a image magnetic charge when an electric charge is close to a surface according to the following mechanism \cite{Qi09,Qi11}.
A surface of topological insulator is described by a massless Dirac Hamiltonian in (2+1) dimensions.
In this system, a parity anomaly arises from a ultra violet divergence, resulting in a Hall current perpendicular to the applied field
\begin{align}
 {j^\mu}= \frac{e^2m}{2h|m|}\sum_{\mu\nu\rho} \epsilon^{\mu\nu\rho}\partial_\nu A_\rho,
\end{align}
where $\Av$ represents an vector potential and $m$ is a topological mass \cite{Redlich84}.
An electric charge emitting a radial electric field outside a topological insulator thus creates a circulating electric current on the surface.
A magnetic field generated by this current is equivalent to the one emitted by a magnetic charge inside  topological insulator. 
Therefore, electric charge in a proximity with a topological insulator is coupled with a mirror magnetic charge (monopole), forming a state called a dyon.
This coupling of electric and magnetic charges is explained also by an effective Lagrangian of a topological insulator in three space dimensions (called a $\theta$ term) \cite{Qi09}
\begin{align}
L_\theta=\frac{\theta}{2\pi}\int d^3x \Ev\cdot\Bv, \label{Ltheta}
\end{align}
where $\theta$ is a constant.
This Lagrangian, derived by integrating out the electrons, indicates that electric field and magnetic field are coupled directly, i.e., there is a magneto-electric effect.
Lagrangian $L_\theta$ is written as a surface integral, and reduces to a Lagrangian which describes a parity anomaly in (2+1) dimensions.

\section{Hedgehog monopole in ferromagnets \label{sec:hhmp}}

\subsection{Gauge field representation of ferromagnetic metals}

In this section, we derive a hedgehog monopole in a metailic ferromagnet in a strongly spin-polarized case.
There are conduction electrons and local spins (magnetization). 
Local spin we consider is a classical vector field, represented by $\Sv({\bm r},t)$, which depends on space coordinate, ${\bm r}$,  and time, $t$.
We use a field (second-quantized) representation, where $c\equiv({c_+},{c_-})^{\rm t}$ 
(lower index $\pm$ denotes spin and t stands for transpose) and $c^\dagger$ represent the annihilation and creation operators for a conduction electron, respectively.
The free part of conduction electron Hamiltonian is 
\begin{align}
H_0=  \intr c^\dagger\left( -\frac{\hbar^2}{2m}\nabla^2 -\eF \right) c , \label{H0}
\end{align}
where $m$ is the electron mass and $\ef$ is the Fermi energy.
In metallic ferromagnets, conduction electrons are spin-polarized by local spin, $\Sv$, due to a coupling of $sd$-type given by 
\begin{align}
H_{\rm sd}=-J \intr \Sv\cdot(c^\dagger \sigmav c),
\end{align}
where  $J$ is a coupling constant, and $(c^\dagger \sigmav c)$ is the spin density of the electron 
($\sigmav=\sigma_x,\sigma_y,\sigma_z$ represents Pauli matrix).
The total Lagrangian of conduction electrons,
defined as 
$L\equiv \intr i\hbar \cdag \dot{c} -H$, where $H=H_0+H_{\rm sd}$, is 
\begin{eqnarray}
L=  \intr \lt[i\hbar \cdag \dot{c} 
-  \left( \frac{\hbar^2}{2m}|\nabla c|^2 -\eF c^\dagger c \right)
+ {\spol}   \evs \cdot (c^\dagger \sigmav c) \rt],
  \label{Le0}
\end{eqnarray}
where $\spol\equiv JS$ and $\nv\equiv \Sv/S$ is a unit vector.
In ferromagnetic metals, spin polarization of conduction electron satisfies $\spol\tau/\hbar\gg1$, where $\tau$ is elastic lifetime of conduction electron.
This is the adiabatic condition for disordered metals\cite{TKS_PR08,Stern92,Popp03}.
In this limit, a local gauge transformation  to choose the electron spin quantization axis along $\Sv(\rv,t)$ at each point is useful \cite{TKS_PR08}.
The deviation from perfect adiabaticity is then described by an SU(2) gauge field, which is small and we treat it perturbatively.
A new electron operator $a\equiv({a_+},{a_-})^{\rm t}$ is defined as
\begin{equation}
c(\rv,t)\equiv U(\rv,t) a(\rv,t),
\end{equation}
where $U$ is a $2\times2$ matrix which we further define as
\begin{equation}
U(\rv,t)\equiv \mv\cdot\sigmav,
\end{equation}
$\mv$ being a real three-component unit vector we will define later.
The matrix satisfies $U^2=1$, i.e., $U(\rv,t)^{-1}=U(\rv,t)$.
A derivative of an operator $c$ reads 
\begin{equation}
\partial_\mu c(\rv,t)=U(\rv,t)(\partial_\mu+U(\rv,t)^{-1}\partial_\mu U(\rv,t))a
  = U(\rv,t)(\partial_\mu+iA_{{\rm s},\mu})a ,
\end{equation}
where a gauge field (represented by a $2\times2$ matrix) is defined as
\begin{equation}
A_{{\rm s},\mu}\equiv -iU(\rv,t)^{-1}\partial_\mu U(\rv,t).
\end{equation}
In terms of spin components, $A_\mu$ is written as \cite{TKS_PR08}
\begin{equation}
A_{{\rm s},\mu}=(\mv\times\partial_\mu \mv)\cdot \sigmav \equiv \sum_{\alpha}A_{{\rm s},\mu}^\alpha \sigma_\alpha  .
\end{equation}
By the above gauge transform, the electron spin is transformed to be
\begin{equation}
U^{-1}\sigmav U =2 \mv(\mv\cdot\sigmav)-\sigmav.\label{sigmatr}
\end{equation}
The aim of our gauge transform is to let this spin to be along $z$-axis, {\it i.e.},
$U^{-1}(\nv \cdot \sigmav)U = \sigma_z$.
This is satisfied if we choose
\begin{equation}
\mv=\left(
\sin\frac{\theta}{2}\cos\phi,\sin\frac{\theta}{2}\sin\phi,\cos\frac{\theta}{2} \right),
\end{equation}
where $(\theta,\phi)$ are the polar coordinates of $\Sv$.
The gauge field is then obtained
in a matrix notation with respect to spin index as
\begin{equation}
\vecth{A_{{\rm s},\mu}^x}{A_{{\rm s},\mu}^y}{A_{{\rm s},\mu}^z}
= \hf 
\vecth{
-\partial_\mu \theta \sin \phi -\sin\theta \cos\phi \partial_\mu \phi }{
\partial_\mu \theta \cos \phi -\sin\theta \sin\phi \partial_\mu \phi }{
  (1-\cos\theta)\partial_\mu \phi  }.\label{Asdef}
\end{equation}

The electron part of Lagrangian is written in terms of $a$-electron as
\begin{eqnarray}
L &=&   \intr \left[
i\hbar \adag \dot{a} -\frac{\hbar^2}{2m}|\nabla a|^2 +\eF \adag a
 -\spol \adag \sigma_z a
  \right.\nonumber\\
 && \left.+i\frac{\hbar^2}{2m}(\adag A_{{\rm s},i} \nabla_i a - (\nabla_i\adag) A_{{\rm s},i} a )
  -\frac{\hbar^2}{2m}(\Av_{\rm s})^2 \adag a -\hbar \adag A_{{\rm s},0} a
 \right].
\end{eqnarray}
In the matrix notation of the spin,
\begin{eqnarray}
L &=& 
 \intr\lt[  (\adag_+,\adag_-) 
\left(
i\hbar \partial_t  +\frac{\hbar^2}{2m}\nabla ^2 +\eF  
-\frac{\hbar^2}{2m}(\Av_{\rm s})^2 
\rt)\lt(
\begin{array}{c}
a_+ \\ a_- \end{array}\rt)
 \rt.  \nnr
&&
\lt. +
  (\adag_+,\adag_-) 
\left( \begin{array}{cc}
-\spol 
+i\frac{\hbar^2}{2m}A_{{\rm s},i}^z {\nablalr}_i
-\hbar A_{{\rm s},0}^z
& 
i\frac{\hbar^2}{2m}A_{{\rm s},i}^+ {\nablalr}_i  -\hbar A_{{\rm s},0}^+ 
\\
i\frac{\hbar^2}{2m}A_{{\rm s},i}^- {\nablalr}_i  -\hbar A_{{\rm s},0}^- 
&
\spol 
-i\frac{\hbar^2}{2m}A_{{\rm s},i}^z {\nablalr}_i
 +\hbar A_{{\rm s},0}^z
\end{array} \rt)
\lt(
\begin{array}{c}
a_+ \\ a_- \end{array}\rt)  \rt],\label{LSU2matrix}
\end{eqnarray}
where 
\begin{eqnarray}
A_{{\rm s},\mu}^\pm &\equiv& A_\mu^x \pm  i A_\mu^y,
\end{eqnarray}
and ${\nablalr}\equiv \nablar-\nablal$ acts only to the field operators.

The adiabatic limit is defined as $\spol\ra\infty$. In this limit, the minority spin electron has infinitely high energy and thus does not exist.
Off-diagonal elements of \Eqref{LSU2matrix} are accordingly neglected, and the system reduces to 
an electron interacting with a U(1) gauge field,  $A_{{\rm s},\mu}^z$, described by a Lagrangian 
($(\Av_{{\rm s}}^x)^2$ and $(\Av_{{\rm s}}^y)^2$  act only as a potential)
\begin{eqnarray}
L_{\rm ad} &=&   \intr \adag_+
\left[ 
i\hbar \partial_t  +\frac{\hbar^2}{2m}\nabla ^2 +\eF  -\spol 
+i\frac{\hbar^2}{2m}A_{{\rm s},i}^z {\nablalr}_i
-\frac{\hbar^2}{2m}(\Av_{\rm s})^2 -\hbar A_{{\rm s},0}^z
\rt]
a_+ .\label{Ladiabatic}
\end{eqnarray}

\subsection{Hedgehog monopole arising from non-adiabaticity}

When $M$ is finite, perpendicular fluctuation represented by $A_{{\rm s},\mu}^\pm$ 
exists, and there is a finite deviation from the U(1) symmetry.
These components appears in the U(1) space as a singular magnetic structure, i.e., a monopole.
This can be shown as follows.
(The notation in this subsection is a relativistic one \cite{Ryder96}, and upper and lower indices have different meanings. See Sec. \ref{sec:relativistic} for details.)
A field strength of the SU(2) gauge fields is
\begin{align}
F_{\rm s}^{\mu\nu}\equiv \sum_{\alpha=x,y,z} F_{\rm s}^{\mu\nu,\alpha}\sigma_\alpha,
\end{align}
where $\alpha$-component is given as
\begin{align}
F_{\rm s}^{\mu\nu,\alpha} 
\equiv
 \partial^\mu A_{\rm s}^{\nu,\alpha}-\partial^\nu A_{\rm s}^{\mu,\alpha} +({\bm A}_{\rm s}^{\mu}\times {\bm A}_{\rm s}^\nu)^\alpha
=
 \partial^\mu A_{\rm s}^{\nu,\alpha}-\partial^\nu A_{\rm s}^{\mu,\alpha} 
+\sum_{\beta\gamma=x,y,z}\epsilon_{\alpha\beta\gamma}  A_{\rm s}^{\mu,\beta} A_{\rm s}^{\nu,\gamma},
\end{align}
where $\epsilon_{\alpha\beta\gamma}$ is the asymmetric tensor in three-dimensions.
By definition, the field strength satisfies the following identity called the Bianchi identity: 
\begin{align}
\epsilon^{\mu\nu\rho\sigma}D_\nu F_{{\rm s},\rho\sigma}
=0,
\end{align}
where $\epsilon_{\mu\nu\rho\sigma}$ is the asymmetric tensor in four-dimensions and 
\begin{align}
(D_\nu F_{{\rm s},\rho\sigma})^\alpha 
\equiv \partial_\nu F_{{\rm s},\rho\sigma}^\alpha
+\sum_{\beta\gamma}\epsilon_{\alpha\beta\gamma}A_{{\rm s},\nu}^\beta F_{{\rm s},\rho\sigma}^\gamma,
\label{bianchi}
\end{align}
represents the covariant derivative of field strength.

When away from the perfect adiabatic limit, $A_{{\rm s},\nu}^x= A_{{\rm s},\nu}^y$ need to be taken account of.
Nevertheless, when the non-adiabaticity is weak, only the $z$ component of the field strength is essential, which reads 
\begin{align}
F_{\rm s}^{\mu\nu,z}= \partial^\mu A_{\rm s}^{\nu,z}
-\partial^\nu A_{\rm s}^{\mu,z} +\Phi^{\mu\nu},
\end{align}
where
\begin{align}
\Phi^{\mu\nu}\equiv(A_{\rm s}^{\mu,x} A_{\rm s}^{\nu,y}-A_{\rm s}^{\nu,x} A_{\rm s}^{\mu,y}),
\end{align}
 is an anomalous field strength representing a trace of the SU(2) gauge field.
In terms of a unit vector  ${\bm n}$, it  reads \cite{TKS_PR08}
\begin{align}
\Phi^{\mu\nu}=\frac{1}{4}{\bm n}\cdot(\partial^\mu{\bm n}\times \partial^\nu{\bm n}).
\end{align}
In the adiabatic limit, $F_{\rm s}^{\mu\nu,\alpha}$ with spin component $\alpha=x,y$ are suppressed and the $z$ component of the Bianchi identity (\ref{bianchi}) reduces to
\begin{align}
\epsilon^{\mu\nu\rho\sigma}\partial_\nu F_{{\rm s},{\rho\sigma,z}}
=\epsilon^{\mu\nu\rho\sigma}\partial_\nu (\partial_\rho A_{{\rm s},\sigma}^z
-\partial_\sigma A_{{\rm s},\rho}^z +\Phi_{\rho\sigma})=0.
\label{bianchiz}
\end{align}
>From a gauge invariance, the effective electric and magnetic fields are defined as
\begin{align}
{E}_{{\rm s},i}&\equiv F_{{\rm s},0i}=-\nabla_i A_{{\rm s},0}^z+\partial_t {A}_{{\rm s},i}^z
=
-\frac{1}{2} \nv \cdot (\dot{\nv} \times \nabla_i \nv)
\nnr
{B}_{{\rm s},i}&=(\nabla\times {\bm A}_{\rm s}^z)_i
=\frac{1}{4}\sum_{jk}\epsilon_{ijk} \nv \cdot (\nabla_j \nv \times \nabla_k \nv).
\label{EandBHH} 
\end{align}
The $\mu=0$ component of \Eqref{bianchiz} then becomes
\begin{align}
-\nabla\cdot\Bv_{\rm s}+\sum_{ijk}\epsilon_{ijk}\partial_i \Phi_{jk}=0,
\end{align}
namely
\begin{align}
\nabla\cdot\Bv_{\rm s}=\rhoh,\label{divBhh1}
\end{align}
where 
\begin{align}
\rhoh \equiv \sum_{ijk} \epsilon_{ijk}\partial_i \Phi_{jk}
=\frac{1}{4}\sum_{ijk}\epsilon_{ijk}
[\partial_i \nv\cdot(\partial_j \nv\times \partial_k \nv)].\label{rhoHdef}
\end{align}
Thus the effective magnetic field has a finite divergence, i.e., a finite monopole density.
Similarly, $\mu=i$ component of \Eqref{bianchiz} reads
\begin{align}
(\nabla\times \Ev_{\rm s})_i + \partial_t\cdot\Bv_{{\rm s},i}
=-j_{{\rm h},i}, \label{Faradayhh1}
\end{align}
where
\begin{align}
j_{{\rm h},i}
 \equiv
\sum_{jk}\epsilon_{ijk}(\partial_t \Phi_{jk}-2\partial_j\Phi_{0k})
=
\frac{3}{4}\sum_{jk}\epsilon_{ijk}
[\dot{\nv}\cdot(\partial_j \nv\times \partial_k \nv)]. \label{jhdef}
\end{align}
Consistency of Eqs. (\ref{divBhh1}) (\ref{Faradayhh1}) is guaranteed by a conservation law for monopole,
\begin{align}
\dot{\rhoh}+\nabla \cdot \jv_{{\rm h}}
=0.
\end{align}
Therefore, ferromagnetic metals having a singular hedgehog spin structure in the adiabatic limit contains monopole.

\subsection{Quantization condition of hedgehog monopole \label{sec:quantizationhhmp}}

An important feature of hedgehog monopole is that its density and current, $\jH$ and $\rhoH$, vanish when the length of local spin is constant.
This is easily seen by noting that a unit vector $\nv$ is described by two independent angles, $\theta$ and $\phi$, and that three vectors $\partial_i \nv$, $\partial_j \nv$ and $\partial_k \nv$ in Eqs. (\ref{rhoHdef}) and   (\ref{jhdef}) cannot be independent.
Nevertheless, the volume integral of the monopole density is finite  due to the surface contribution if the local spin has a hedgehog structure shown in Fig.~\ref{fig:hedgehog}.
Let us define a normalized spin gauge field including a copling constant $\frac{g}{2\pi}$, i.e., as
(upper suffix N means north)
\begin{align}
{A}_{\mu}^{\rm N}\equiv \frac{g}{4\pi}(1-\cos\theta)\partial_\mu \phi .\label{Astilde}
\end{align}
The effective magnetic field of ${\Av}^{\rm N}$ reads
\begin{align}
{{B}^{\rm N}}_{i} \equiv (\nabla\times{\Av}_{\mu}^{\rm N})_i
=\frac{g}{8\pi}\sum_{jk}\epsilon_{ijk} \nv \cdot (\nabla_j \nv \times \nabla_k \nv).
\label{BHHtilde} 
\end{align}
The total magnetic flux is  then 
\begin{align}
\int d\Sv\cdot {{\Bv}}^{{\rm N}}
={gn},
\end{align} 
meaning that there are $n$ monopoles with a charge $g$.
Here we used a relation
\begin{align}
\sum_{ijk} \int dS_i \epsilon_{ijk} \nv \cdot (\nabla_j \nv \times \nabla_k \nv)
=2 \sum_{ijk} \int dS_i \epsilon_{ijk} \sin\theta (\nabla_j \theta) (\nabla_k \phi)
=8\pi n, \label{windingnumber}
\end{align}
where $n$ is an integer.
This is understood by noticing that $\sin\theta d\theta d\phi$ is an element of are of a sphere spanned by $\theta$ and $\phi$, and thus an integral 
$\sum_{ijk} \int dS_i \epsilon_{ijk} \sin\theta (\nabla_j \theta) (\nabla_k \phi)$ is a solid angle of a sphere ($4\pi$) times a winding number of spin structure, an integer $n$.
This relation is a result of a fact that the total solid angle subtended by a spin structure is $4\pi n$, or equivalently a fact that the left-hand side of \Eqref{windingnumber} is a winding number multiplied by $8\pi$. 
The hedgehog monopole from the spin structure is thus a topological object having vanishing local densities. 
The one shown in Fig. \ref{fig:hedgehog} has a topological number of  $n=1$.

A quantization of monopole charge $g$ is discussed by requiring that a gauge field covering the whole space without singularity is constructed by patching together locally defined gauge fields.
In fact, \Eqref{Astilde} is not defined at $\theta=\pi$ (south pole), since $\partial \phi$ can not be defined there.
We can define a gauge field regular at the south pole as
\begin{align}
{A}_{\mu}^{\rm S}= -\frac{g}{4\pi}(1+\cos\theta)\partial_\mu \phi .\label{ASouth}
\end{align}
This field has a singularity at the north pole $\theta=0$, but represents the same magnetic field as \Eqref{BHHtilde}.
We can define therefore only a gauge field with a singularity, if we try to describe the whole space by a single gauge field. 
The singularity is regarded as a Dirac's string.
Instead, we can cover the whole space by patching two gauge fields, ${A}_{\mu}^{\rm S}$ and \Eqref{Astilde}, which we call ${A}_{\mu}^{\rm N}$.
They are related by a gauge transformation
\begin{align}
{A}_{\mu}^{\rm N}={A}_{\mu}^{\rm S} +i\frac{\hbar}{e} \Theta^{-1}\partial_\mu \Theta,
\end{align}
where 
\begin{align}
\Theta \equiv e^{-i\frac{eg}{h}\phi},
\end{align}
is a gauge transform function.
This function must be single valued, i.e., is invariant under $\phi\ra\phi+2\pi$.
Thus a condition 
\begin{align}
{eg}=2\pi{h},\label{diracquantization2}
\end{align}
is imposed, which is a Dirac's quantization condition (\Eqref{diracquantization}).

Mechanism of generation of a hedgehog monopole is essentially the same as a GUT monopole, namely, a symmetry breaking of a non-Abelian gauge field (SU(2) for a hedgehog and SU(5) for GUT).
There is, however, a difference in the two models.
A GUT monopole is a composite object of a gauge field and a Higgs field, and the monopole solution is solved in the same footing as a Higgs field.
In a case of hedgehog monopole, in contrast, local spin structure is treated as a background field which is treated as not to be affected by gauge field.

The effective magnetic field of a hedgehog monopole configuration, \Eqref{EandBHH}, is sometimes called the spin Berry's phase, and the electric field is known as the spin motive force \cite{Stern92}.
The magnetic field has been observed for example in the anomalous Hall effect  \cite{Nagaosa10}, and the electric field has been observed by inducing magnetization dynamics such as domain wall motion \cite{Yang09}.
These experiments do not, however, mean existence of hedgehog monopole, since as explained in \S\ref{sec:topologicalobjects}.

\subsection{Hedgehog monopole in weak exchange coupling regime \label{sec:hhmpweakcoupling}}

Until very recently, a hedgehog monopole has been discussed exclusively in the adiabatic (strong sd coupling) regime.
As we have discussed in the earlier sections, a symmetry breaking of spin SU(2) space to a U(1) space of  electromagnetism is clearly defined in this regime.
However, hedgehog monopole is not an object restricted to adiabatic regime.
In fact, introduction of an exchange interaction to rotationally invariant spin results in a symmetry breaking 
even if the interaction is weak.
In this section, we will demonstrate that the hedgehog monopole emerges even if the exchange coupling $J$ is small following the analysis in Ref. \cite{Takeuchi11a}.
We cannot approach monopoles in this regime by a standard gauge field argument. 
Here we apply instead a novel method pointed out in Ref. \cite{Takeuchi12,Takeuchi11a} based on a transport calculation.
In a transport method,  effective electromagnetic fields are calculated by evaluating electric charge density and current density induced by local spin structure by use of Keldysh Green's functions.
A monopole field is then identified by deriving the Maxwell's equation for the effective fields.

We consider a disordered metal, and take account of the spin-independent impurity scattering represented as
\begin{align}
H_{\rm i} = \int{d^3 r} c^\dagger \vi c,
\end{align}
$\vi$ being the impurity potential.
In the following calculation, the impurities are approximated as random point scatterers and the averaging is carried out as 
\begin{align}
\langle{v_{\rm i}({\bm r}) v_{\rm i}({\bm r}')}\rangle_{\rm i} = n_{\rm i} u_{\rm i}^2 \delta^3({\bm r}-{\bm r}'),
\end{align}
where $n_{\rm i}$ and $u_{\rm i}$ are the impurity concentration and the strength of the scattering, respectively ~\cite{Rammer07}.
The impurities give rise to an elastic lifetime for the electron, $\tau$, which is calculated as 
\begin{align}
\tau = \frac{\hbar}{2\pi n_{\rm i}u_{\rm i}^2 \nu},
\end{align}
 ($\nu$ is the density of states per volume).
The total Hamiltonian discussed in this section is  
\begin{align}
H = H_0 +H_{\rm i} +H_{\rm sd}.
\end{align}
The electric charge density is 
\begin{align}
\rho(\rv,t) = -e \tr \langle{c^\dagger(\rv,t) c(\rv,t)}\rangle,
\end{align}
 where the bracket represents the quantum expectation value and $\tr$ is a trace over spin indices, and  electric current density is given as
\begin{equation}
j_i(\rv,t)
=
\frac{e \hbar^2}{2m} (\nabla_{\rv} -\nabla_{\rv'})_i \tr
G^<(\rv,t; \rv',t) \big|_{\rv'=\rv},
\end{equation}
where 
\begin{align}
G_{s,s'}^<(\rv,t; \rv',t') \equiv \frac{i}{ \hbar} \langle{c^\dagger_{s'}(\rv',t') c_s(\rv,t)}\rangle,
\end{align}
 ($s$ and $s'$ are spin indices) is the lesser component of the non-equilibrium Green's function \cite{Haug07}.
This Green's function defined on the Keldysh contour C satisfies the Dyson's equation,
\begin{align}
G_{s s'}({\bm r},t;{\bm r}',t')
=
&\delta_{s, s'}
g_s({\bm r},t;{\bm r}',t')
\notag
\\
&+\int{d^3r''} \int_{\rm C}{dt''}
g_s({\bm r},t;{\bm r}'',t'')
\notag
\\
&\times
[
\delta_{s, s''} v_{\rm i}({\bm r}'')
-J \Sv({\bm r}'',t'') \cdot {\bm \sigma}_{s s''}
]
\notag
\\
&\times
G_{s'' s'}({\bm r}'',t'';{\bm r}',t'),
\end{align}
where $G_{s s'}({\bm r},t;{\bm r}',t') \equiv -(i/\hbar) \langle{{\rm T_C} [c_s({\bm r},t) c^\dagger_{s'}({\bm r}',t')]}\rangle$ 
(${\rm T_C}$ is the path-ordering operator on C)
and $g$ denotes free Green's function.
This equation is solved by iteration.
Here, we assume slowly varying magnetization profile $\Sv_{\qv,\Omega}$ in space and time: the spatially smooth magnetization structure compared to the electron mean free path $\ell$, $q \ell \ll 1$ ($q$ is a wave number of magnetization texture), and the sufficiently slow dynamics of magnetization, $\Omega \tau \ll 1$ ($\Omega$ is a frequency of magnetization dynamics).

\begin{figure}[tb]
\begin{center}
\includegraphics[width=0.3 \hsize]{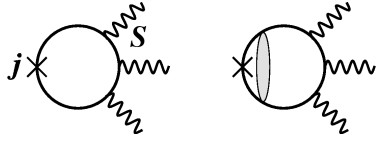}
\end{center}
\caption{
Diagrammatic representation of the electric current, $\jv$, induced by sd interaction with local spin, $\Sv$.
Solid lines represent the conduction electron's Green's function and wavy lines are the interection with $\Sv$.
The first diagram is a contribution of an effective electric and magnetic fields local in space, and the second diagram, containing a diffusion ladder (vertex corrections) denoted by the gray shaded oval, results in a diffusive current (a gradient of electric charge density).
}
\label{fig;D-j_sd}
\end{figure}

To see a hedgehog monopole, it is enough to discuss the electric current to the third-order in the exchange coupling, $J$.
This contribution is diagrammatically shown in Fig.~\ref{fig;D-j_sd} and is calculated as
\begin{align}
j_i(\rv,t)
=
&\frac{eJ^3}{\pi mV}
\sum_{\kv,\qv,\qv',\Qv} \sum_{\omega,\omega',\Omega}
e^{-i\Qv\cdot\rv +i\Omega t}
\Sv_{\qv,\omega} \cdot (\Sv_{\qv',\omega'} \times \Sv_{\Qv-\qv-\qv',\Omega-\omega-\omega'})
\notag
\\
&\times \bigg[
\frac{i\hbar^3}{30m} q_i (\Qv \cdot \qv') \Im (\ga_\kv)^4
-2\tau^2 \omega q'_i |\ga_\kv|^2
\bigg]
-D \nabla_i \rho(\rv,t),\label{jtochu}
\end{align}
where 
\begin{align}
\ga_\kv = \frac{1}{\ef -\frac{\hbar^2 \kv^2 }{ 2m} -\frac{i\hbar }{ 2\tau}},
\end{align}
 is the advanced Green's function ($\ef$ is the Fermi energy), $\Im$ means taking an imaginary component, $V$ is system volume, and $D = 2\ef\tau / 3m$ denotes a diffusion constant.
The last term is a diffusive contribution arising from vertex corrections shown in the right diagram of Fig.~\ref{fig;D-j_sd} and  electric charge density, $\rho$, is calculated as
\begin{equation}
\rho
=
\frac{4e\nu J^3 \tau^4}{\hbar m}
\nabla_i
\average{\Sv \cdot (\dot{\Sv} \times \nabla_i \Sv)}_{\rm D},
\end{equation}
where $\langle{\cdots}\rangle_{\rm D}$ is an average including  electron diffusion, which satisfies
\begin{align}
(-D\nabla^2 +\partial_t) \langle{F(\rv,t)}\rangle_{\rm D}
=
\frac{1}{\tau} F(\rv,t), \label{diffusioneq}
\end{align}
 ($F$ is an arbitrary function depending on space $\rv$ and time $t$).

Summing over the wave vectors in \Eqref{jtochu}, we obtain the electric current,
\begin{equation}
j_i
=
-\frac{e\hbar^3\nu J^3}{960m^2\ef^3}
\epsilon_{ijk} \epsilon_{klm} \nabla_j \big[ \Sv \cdot (\nabla_l \Sv \times \nabla_m \Sv) \big]
-\frac{4e\nu J^3 \tau^3}{\hbar m}
\Sv \cdot (\dot{\Sv} \times \nabla_i \Sv)
-D \nabla_i \rho.
\end{equation}
As was pointed our in Ref. \cite{Takeuchi11a},
the effective electric and magnetic fields ($\Ev_{{\rm h}}$ and $\Bv_{{\rm h}}$) are read from the above result by comparing it with a general expression, 
\begin{align}
\jv = (1 / \mu) \nabla\times \Bv_{{\rm h}} +\sigma_{\rm c} \Ev_{{\rm h}} -D \nabla \rho,
\label{Jgeneral}
\end{align}
 where $\mu$ is  magnetic permeability, 
\begin{align}
\sigmac\equiv e^2n\tau / m=\frac{e^2}{3}\frac{\nu}{V}\lt(\frac{\hbar \kf}{m}\rt)^2\tau,
\end{align}
 is  electric conductivity.
The result is
\begin{align}
E_{{\rm h}, i}
&=
-2\gamma_{\rm h} 
\Sv \cdot (\dot{\Sv} \times \nabla_i \Sv),
\\
B_{{\rm h}, i}
&= \gamma_{\rm h} 
\epsilon_{ijk} \Sv \cdot (\nabla_j \Sv \times \nabla_k \Sv),
\label{eq;fields}
\end{align}
where 
\begin{align}
\gamma_{\rm h} \equiv 6\hbar \frac{m J^3 \tau^2}{\hbar^4 \kf^2},
\end{align}
 and 
we defined  magnetic permeability as
\begin{align}
\frac{1}{\mu}= -\frac{1}{45\cdot2^5}\frac{e^2}{\hbar^2\kf^4}\frac{\nu}{V}\lt(\frac{\hbar}{\tau}\rt)^2.
\end{align}
Obviously, these effective fields satisfy the Faraday's law and the Gauss's law with  magnetic monopole, \begin{align}
\nabla\times \Ev_{{\rm h}} +\dot{\bm B}_{\rm h} &= -\jHv \nnr
\nabla\cdot \Bv_{{\rm h}} &= \rhoH,
\end{align}
where magnetic monopole contributions are
\begin{align}
j_{{\rm h}, i}
&=
-3\gamma_{\rm h}
\epsilon_{ijk} \dot{\Sv} \cdot (\nabla_j \Sv \times \nabla_k \Sv),
\\
\rhoH
&= \gamma_{\rm h}
\epsilon_{ijk} \nabla_i \Sv \cdot (\nabla_j \Sv \times \nabla_k \Sv).
\label{eq;hedgehog monopole}
\end{align}
There is therefore hedgehog monopole in the weak sd coupling case, too, although the coefficients in 
 \Eqref{eq;hedgehog monopole} differ from the strong coupling limit.
It is notable that the structure of the electromagnetism, i.e., U(1) gauge field, (and further, that with monopole in the present case) is embedded in the electron transport phenomena.  

We note here that 
the definition of $\Ev_{\rm h}$, $\Bv_{\rm h}$, $\mu$ according to \Eqref{Jgeneral} and the dielectric constant $\epsilon$ according to the Gauss's law has arbitrariness.
In fact, the condition imposed by the transport properties ($\jv$ and $\rho$) is not sufficient to fix the two effective fields uniquely, and an additional condition seems to be required.
Nevertheless, existence of monopole holds true; both of the monopole current and density cannot be deleted at the same time by redefining the fields. 
Further, as noted in Ref. \cite{Takeuchi11a}, the product $\epsilon\mu$ is invariant;
a physical quantity of the velocity of the effective topological electromagnetic field is uniquely given as
\begin{equation}
v_{\rm top}
\equiv
\frac{1}{\sqrt{\epsilon\mu}}
=
\frac{1}{8\sqrt{30}} \frac{\hbar}{\kf\ef\tau^2}.
\end{equation}
If elastic mean free path, $\ell\equiv \frac{\hbar \kf \tau}{m}$, is 10\AA, $\ef\tau/\hbar=3.3$ by choosing $\kf^{-1}=1.5$\AA,
the speed of the topological electromagnetic wave is thus rather large, about 900m/s.  
The topological electromagnetic field may be useful to transport spin information in a different manner from magnon transports.

\section{Spin damping monopole \label{sec:sdmp}}

As we have seen, the hedgehog monopole is a topological object and has locally vanishing density and current only.
The hedgehog monopole hence does not locally coupled to the electromagnetism.  
Very recently, a novel monopole in magnets with locally finite density and current density was discovered by Takeuchi et al \cite{Takeuchi12}.
Such a monopole creates a rotational electric field via the Amp\`ere's law (Fig. \ref{fig:ampere}), and thus it acts as an anomalous angular momentum source which induces rotational motion of electric charge.
To realize such a monopole, Takeuchi et al. included the spin-orbit interaction.

\begin{figure}
\begin{center}
\includegraphics[scale=0.65]{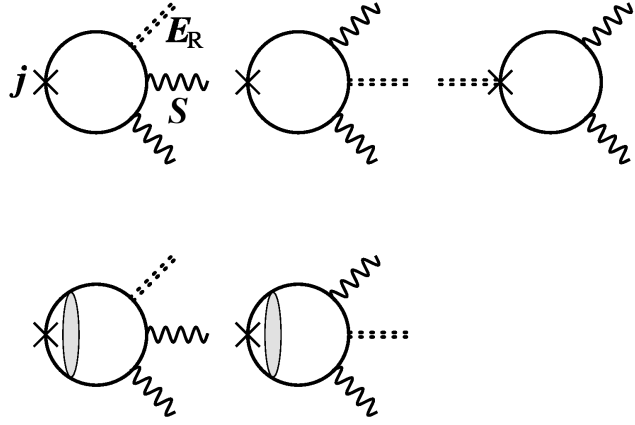}
\end{center}
\caption{
Diagrammatic representations of electric current pumped by magnetization dynamics and the  Rashba interaction.
Solid lines represent the conducting electron Green's functions, and double dashed and wavy lines represent the Rashba spin-orbit interaction (${\bm E}_{\rm R}$) and the interaction with localized spin (${\bm S}$), respectively.
The first three contributions correspond to effective electromagnetic fields, and the last two contributions, containing diffusion ladders (vertex corrections) denoted by the gray shaded ovals.
}
\label{fig;j_rashba}
\end{figure}

\begin{figure}
\begin{center}
\includegraphics[scale=0.65]{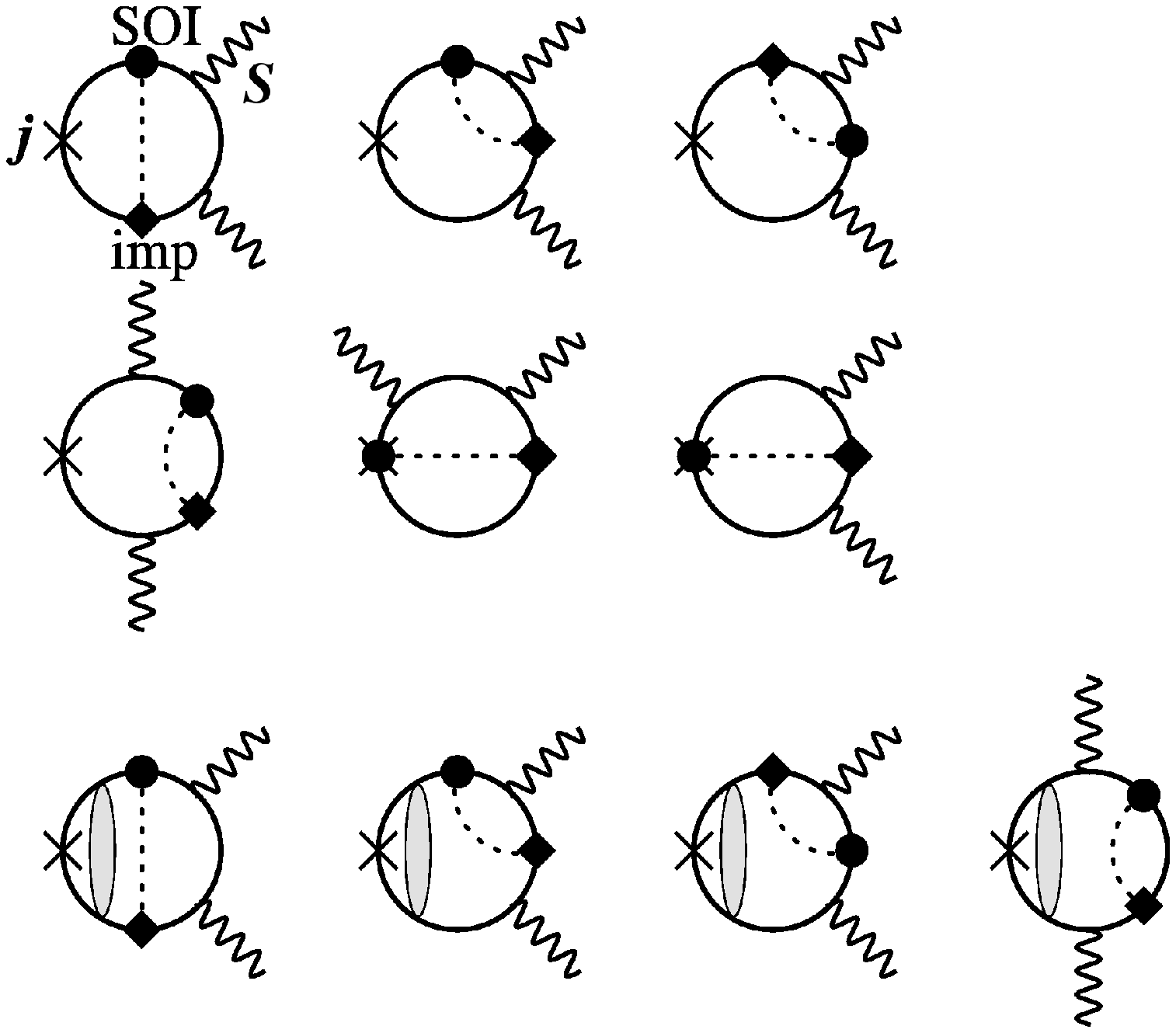}
\end{center}
\caption{
Contribution to the electric current driven by a precession of localized spin and random impurity-induced spin-orbit interaction.
Filled circles are spin-orbit interactions arising from random impurities (SOI), filled diamonds are nonmagnetic impurity scatterings (imp), and dotted lines linking filled circles to filled diamonds represent impurity average.
}
\label{fig;j_imp}
\end{figure}

The spin-orbit interaction exists in any elements including magnetic ones, and is particularly strong in heavy elements such as platinum and gold, and at the interfaces in junctions where the inversion symmetry is broken~\cite{Meier_nature07}.
Two types of the spin-orbit interaction were thus considered in Ref. \cite{Takeuchi12}.
The first is the one from a uniform field, ${\bm E}_{\rm R}$, namely the Rashba interaction~\cite{Rashba60}.
Such a field is realized at the interfaces and surfaces.
The Rashba interaction in metallic films recently turned out to be particularly useful for current-driven magnetization switching \cite{Obata08,Miron10}. 
The second is the one from a random potential, $v_{\rm i}$, induced by heavy impurities.
This random heavy impurity model would also simulate the effect of spin-orbit interaction in heavy pure metals.
The total spin-orbit interaction thus reads
\begin{equation}
H_{\rm so}
=
-\frac{1}{\hbar}
\intr c^\dagger \lt[( \lambda_{\rm R}{\bm E}_{\rm R}
-\lambda_{\rm i} {\bm \nabla} v_{\rm i} )
\cdot ({\bm p} \times {\bm \sigma})\rt] c,
\end{equation}
where ${\bm p}$ is the electron's momentum and $\lambda$ is a spin-orbit coupling constant (the subscript $\rm R$ and $\rm i$ characterize Rashba and impurity-induced ones, respectively).
The interaction with the magnetization is described by 
 $H_{\rm sd}$.
The Hamiltonian of the present system is, therefore, given as 
\begin{align}
H = H_0+H_{\rm i} +H_{\rm sd} +H_{\rm so}.
\end{align}
The electric current, ${\bm j}$, generated in the system by magnetization dynamics 
is calculated by evaluating a quantum field theoretical expectation value of electron velocity operator,
\begin{align}
\hat{\bm v} = -(i\hbar / m) {\bm \nabla} +(1/\hbar) (\lambda_{\rm R} {\bm E}_{\rm R} -\lambda_{\rm i} {\bm \nabla} v_{\rm i}) \times {\bm \sigma}.
\end{align}
The electric current thus reads
\begin{align}
{\bm j}({\bm r},t)
=
&e \, {\rm tr}
\lt[
\lt(
\frac{\hbar^2}{2m} ({\bm \nabla}_{\bm r} -{\bm \nabla}_{{\bm r}'})
+i [\lambda_{\rm R} {\bm E}_{\rm R}
-\lambda_{\rm i} {\bm \nabla} v_{\rm i}({\bm r})] \times {\bm \sigma}
\rt)
G^<({\bm r},t;{\bm r}',t)
\rt]_{{\bm r}'={\bm r}}.
\end{align}
This expectation value is evaluated by solving the following Dyson's equation,
\begin{align}
G_{s s'}({\bm r},t;{\bm r}',t')
=
&\delta_{s, s'}
g_s({\bm r},t;{\bm r}',t')
\notag
\\
&+\int{d^3r''} \int_{\rm C}{dt''}
g_s({\bm r},t;{\bm r}'',t'')
\notag
\\
&\times
(
\delta_{s, s''} v_{\rm i}({\bm r}'')
-J \Sv({\bm r}'',t'') \cdot {\bm \sigma}_{s s''}
\notag
\\
&+i \{
[\lambda_{\rm R} {\bm E}_{\rm R}
-\lambda_{\rm i} {\bm \nabla} v_{\rm i}({\bm r}'')]
\times {\bm \nabla}_{{\bm r}''}
\}
\cdot {\bm \sigma}_{s s''}
)
\notag
\\
&\times
G_{s'' s'}({\bm r}'',t'';{\bm r}',t').
\end{align}
This equation is solved treating $\lambda$ and $J$ perturabatively to the linear and second orders, respectively.
Contributions to current is represented by Feynman diagrams in Figs. \ref{fig;j_rashba} and \ref{fig;j_imp}.


\begin{figure}
\includegraphics[scale=0.3]{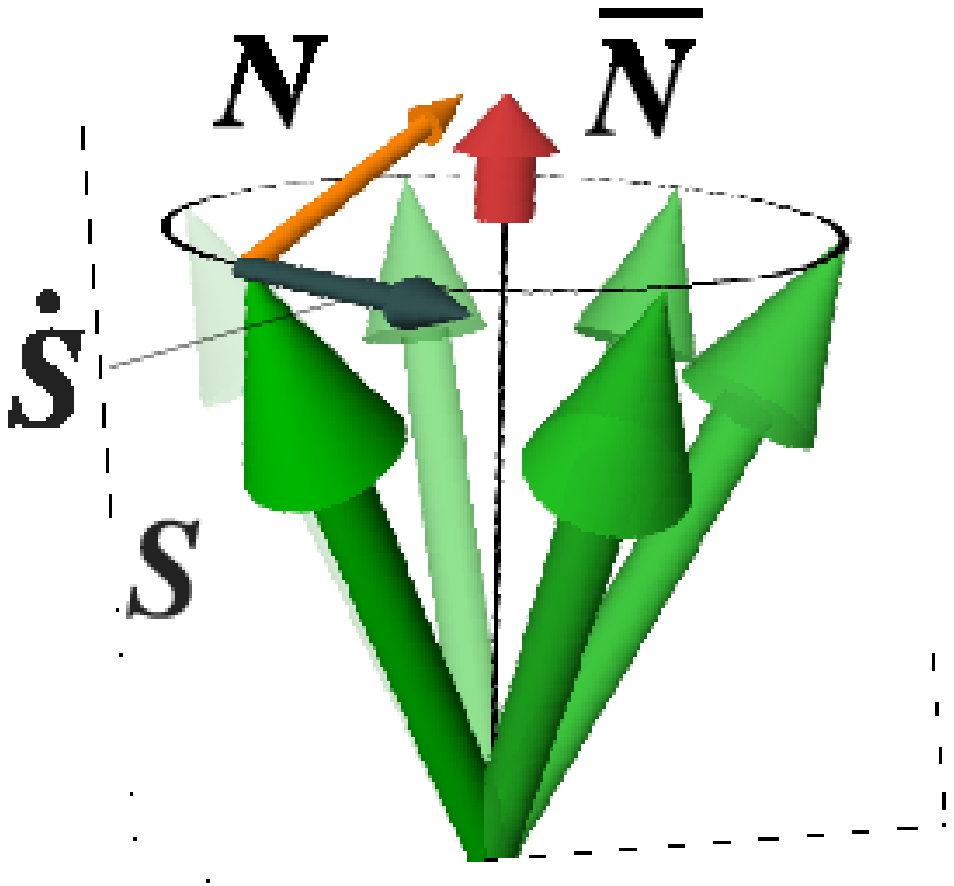}
\includegraphics[width=0.3\hsize]{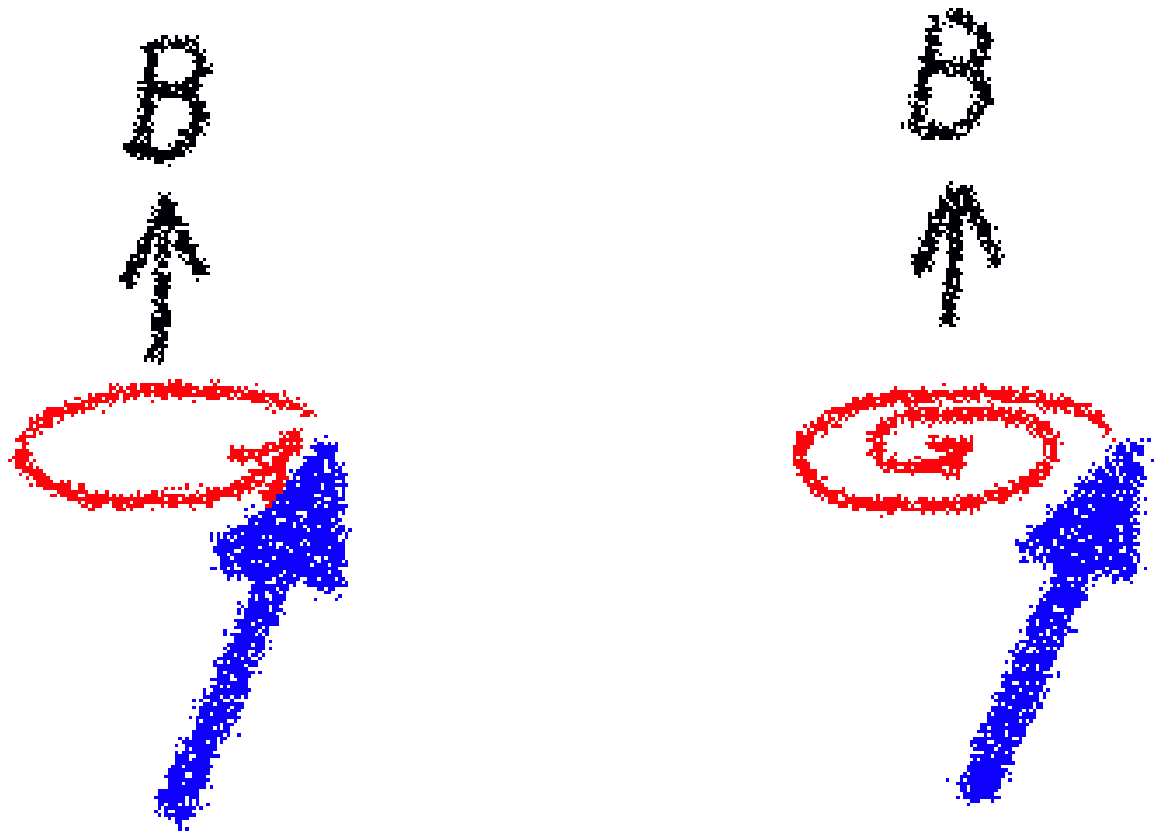}
\caption{
Left: Schematic illustration of damping of a precessing local spin $\bm S$.
A vector product of a time derivative of spin, $\dot{\bm S}$, and $\Sv$ is a vector $\Nv\equiv \Sv\times \dot{\Sv}$.
Its average is $\overline{\Nv}$, which is along the precession axis, and it represents a time-averaged dissipated spin magnitude.
Right: Damping torque, $\Nv$, results in a damping of spin precession when an external magnetic field $\Bv$ is applied.
}
\label{fig:damping}
\end{figure}

We consider sufficiently slow dynamics of magnetization, namely $\Omega \tau \ll 1$ ($\Omega$ is a frequency of magnetization dynamics), and assume that the magnetization structure varies smoothly in the space compared to the electron mean free path $\ell$, i.e., $q \ell \ll 1$ ($q$ is a wave number of magnetization profile).
The leading contribution in this case turns out to be
\begin{align}
{\bm j}({\bm r},t)
=
&-\frac{eJ^2}{2\pi V}
\sum_{{\bm k}, {\bm k}', {\bm q}_1,{\bm q}_2} \sum_{\omega, \Omega_1, \Omega_2}
e^{-i({\bm q}_1+{\bm q}_2)\cdot{\bm r} +i(\Omega_1+\Omega_2)t}
\Omega_1
\notag
\\
&\times
(\Sv_{{\bm q}_1,\Omega_1} \times \Sv_{{\bm q}_2,\Omega_2}) \times
[
\frac{i \lambda_{\rm R} \tau}{\hbar}
{\bm E}_{\rm R}
|g^{\rm a}_{{\bm k}}|^2
\notag
\\
&+\frac{4 \hbar^2 \lambda_{\rm i}}{3\pi \nu \tau^2}
({\bm q}_1 +{\bm q}_2)
\varepsilon_{\bm k}
|g^{\rm a}_{{\bm k}}|^2
|g^{\rm a}_{{\bm k}'}|^4
]
\notag
\\
&-D {\bm \nabla} \rho({\bm r},t).
\label{calculation}
\end{align}
The last term is the diffusive contribution arising from the vertex corrections, where the electric charge density $\rho$ is
\begin{equation}
\rho=
\frac{4e\nu \lambda_{\rm R} J^2 \tau^3}{\hbar^2}
{\bm \nabla} \cdot \langle{ {\bm E}_{\rm R} \times (\Sv \times \dot\Sv) }\rangle_{\rm D}.
\end{equation}
Here $\langle{\cdots}\rangle_{\rm D}$ represents the average including the electron diffusion, defined in \Eqref{diffusioneq}.
Summing over the wave vectors and frequencies in Eq.~(\ref{calculation}), the electric current is obtained as  
\begin{align}
{\bm j}
=
&-\frac{16e\nu \lambda_{\rm i} J^2 \ef \tau^2}{3\hbar^2}
{\bm \nabla} \times (\Sv \times \dot\Sv)
\notag
\\
&-\frac{4e\nu \lambda_{\rm R} J^2 \tau^2}{\hbar^2}
{\bm E}_{\rm R} \times (\Sv \times \dot\Sv)
-D {\bm \nabla} \rho.
\end{align}
This result is rewritten by use of effective electric and magnetic fields, ${\bm E}_{\rm s}$ and ${\bm B}_{\rm s}$, as
\begin{equation}
{\bm j} =
\frac{1}{\mu} {\bm \nabla} \times {\bm B}_{\rm s}
+\sigma_{\rm c} {\bm E}_{\rm s}
-D {\bm \nabla} \rho,
\label{jresult}
\end{equation}
where the effective fields are defined as
\begin{align}
{\bm E}_{\rm s}
&\equiv
-\alpha_{\rm R}
{\bm E}_{\rm R} \times {\bm N},
\notag
\\
{\bm B}_{\rm s}
&\equiv
-\beta_{\rm i}
{\bm N}.
\label{EBdef}
\end{align}
Here 
\begin{align}
{\bm N} \equiv \Sv \times \dot\Sv,
\end{align}
is a vector representing the spin damping torque
(see \S\ref{sec:damping}) (Fig.~\ref{fig:damping})
~\cite{Chikazumi97,TKS_PR08}.
Coefficients $\alpha_{\rm R}$ and $\beta_{\rm i}$ are
\begin{align}
\alpha_{\rm R} & \equiv \frac{4e\nu \lambda_{\rm R} J^2 \tau^2}{\sigma_{\rm c} \hbar^2} \nnr
\beta_{\rm i}  & \equiv \frac{16e\nu \mu \lambda_{\rm i} J^2 \ef \tau^2}{3\hbar^2}.
\end{align}
The effective fields calculated here are the ones acting on the electronic spin in the same manner as the effective fields from the hedgehog monopole.
Clearly, the fields [Eq.~(\ref{EBdef})] do not satisfy the Faraday's law and the Gauss's law of the conventional electromagnetism, but the ones with monopole contribution,
\begin{align}
{\bm \nabla} \times {\bm E}_{\rm s} +\dot{\bm B}_{\rm s} &= -{\bm j}_{\rm m}, \nonumber\\
{\bm \nabla} \cdot {\bm B}_{\rm s} &= \rho_{\rm m},
\label{AmperedivB}
\end{align}
where the monopole current and monopole density read
\begin{equation}
{\bm j}_{\rm m} =
\alpha_{\rm R} {\bm \nabla} \times ({\bm E}_{\rm R} \times {\bm N})
+\beta_{\rm i} \dot{\bm N},
\label{jmdef}
\end{equation}
and
\begin{equation}
\rho_{\rm m} =
-\beta_{\rm i} {\bm \nabla} \cdot {\bm N}.
\label{rhomdef}
\end{equation}

We have thus proved that a monopole emerges when spin damping occurs, namely we have a spin damping monopole.
The spin damping monopole is a composite object made from a magnetization configuration in the same manner as the hedgehog monopole. 
The monopole satisfies the conservation law, $\dot{\rho}_{\rm m} +{\bm \nabla} \cdot {\bm j}_{\rm m} = 0$.

\subsection{Remarks on uniqueness of effective fields}

Equation~(\ref{AmperedivB}) apparently contains an arbitrariness.
In fact, one may think that a transformation 
${\bm E}_{\rm s} \rightarrow {\bm E}_{\rm s} +(1/\sigma_{\rm c}) {\bm \nabla} \times {\bm C}$ and 
${\bm B}_{\rm s} \rightarrow {\bm B}_{\rm s} -\mu {\bm C}$, where ${\bm C}$ is an arbitrary vector field, 
is possible without changing Eq.~(\ref{AmperedivB}).
However, such a transform is not allowed because of a gauge invariance in the original space with a higher symmetry, as is known in the case of hedgehog monopole, where an SU(2) gauge invariance forbids this arbitrariness.

Validness of our definition of ${\bm E}_{\rm s}$ and ${\bm B}_{\rm s}$ in \Eqref{EBdef} is supported by the following argument.
In \Eqref{jresult}, electron diffusion ($D {\bm \nabla} \rho$) induces the effective electric and magnetic polarizations ${\bm P}_{\rm s}$ and ${\bm M}_{\rm s}$, respectively.
By using the relation 
\begin{align}
(-D{\bm \nabla}^2 +\partial_t)\rho = -\sigma_{\rm c} {\bm \nabla} \cdot {\bm E}_{\rm s},
\end{align}
the electric current (eq.~(\ref{jresult})) is described as the rotation of the magnetic field and the time derivative of the electric field, 
\begin{align}
{\bm j} = {\bm \nabla} \times {\bm H}_{\rm s} -\dot{\bm D}_{\rm s},
\end{align} 
where ${\bm H}_{\rm s}$ and ${\bm D}_{\rm s}$ are the fields defined as 
\begin{align}
{\bm H}_{\rm s} & \equiv \frac{1}{\mu} {\bm B}_{\rm s} -{\bm M}_{\rm s} \nnr
{\bm D}_{\rm s} & \equiv \varepsilon {\bm E}_{\rm s} +{\bm P}_{\rm s}, \label{HDdefs}
\end{align}
with $\varepsilon = -\sigma_{\rm c} \tau$ being the permittivity, respectively.
In \Eqref{HDdefs}, ${\bm E}_{\rm s}$ and ${\bm B}_{\rm s}$ are naturally identified with the electric and magnetic fields, respectively.
By use of these fields, two of the Maxwell's equations become
\begin{align}
\nabla\times{\bm D}_{\rm s} +\dot{ {\bm H}}_{\rm s} & = {\bm j}_{\rm m}' \nnr
\nabla\cdot {\bm H}_{\rm s} & = {\rho}_{\rm m}.
\end{align}
Here $ {\bm j}_{\rm m}'$ is redefined monopole current, which is  non-local. 


\begin{figure}
\includegraphics[width=0.5\hsize]{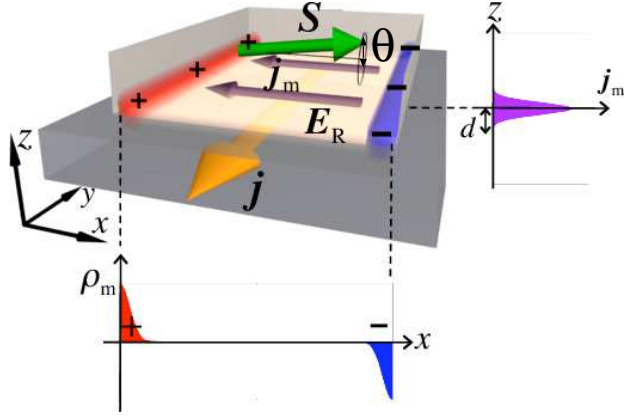}
\caption{
Schematic illustration of monopole pumping and detection in a thin ferromagnetic film attached on a non-magnetic layer.
Magnetization ($\bm S$) precession is induced by applying an oscillating magnetic field.
The amplitude of presession is represented by an angle $\theta$.
The Rashba field, ${\bm E}_{\rm R}$, exists at the interface and creates the monopole current, ${\bm j}_{\rm m}$, near the interface.
The width of the monopole current distribution, $d$,  is comparable to the decay length of the magnetization at the interface.
The monopole current induces an electric current, ${\bm j}$, via the Amp\`ere's law at the interface.
The impurity spin-orbit interaction directly induces positive ($+$) and negative ($-$) monopole charge distribution, $\rho_{\rm m}$, at the two edges.
}
\label{fig;system}
\end{figure}

\subsection{Spin damping monopole generation in a ferro-normal junction}

The spin damping monopole is unique since it does not require a particular non-coplanar spin structure like a hedgehog, and so it exists quite generally in magnetic systems.
The simplest candidate for creating the monopole would be a thin ferromagnetic film put on a non-magnetic insulator or metal shown in Fig.~\ref{fig;system}.
We choose the $z$ axis perpendicular to the film.
The Rashba-type spin-orbit field would then arise at the interface along the $z$ direction~\cite{Meier_nature07}.
We excite the precession of the uniform magnetization by applying the alternating magnetic field in the $yz$ plane in the presence of static field along the $x$ axis (ferromagnetic resonance~\cite{Chikazumi97}).
The precession results in the spin damping vector with a finite time average, $\overline{\bm N}$, along $x$ direction.
In the present case with the uniform magnetization, spatial derivatives in Eqs.~(\ref{jmdef}) and (\ref{rhomdef}) arise at the interface and at the edges, where the magnetization vanishes.
The Rashba interaction contributes to the DC monopole current at the interface as
\begin{align}
\overline{j_{{\rm m}, x}^{\rm R}}
=
-\alpha_{\rm R} E_{\rm R} \frac{\partial \overline{N}}{\partial z}
\simeq
-\frac{\alpha_{\rm R}}{d} E_{\rm R} \overline{N},
\end{align}
where $d$ is the spatial scale of the magnetization decay at the interface.
The monopole current driven by the random spin-orbit impurities, on the other hand, vanishes when time-averaged.
The total DC monopole current thus reads
\begin{align}
\overline{ {\bm j}_{\rm m} }
=
-{\bf e}_x \frac{\alpha_{\rm R}}{d} E_{\rm R} \overline{N},
\end{align}
(${\bf e}_x$ represents the unit vector along the $x$ direction).
This monopole current at the interface generates the electromotive force along the $y$ direction via the Amp\`ere's law for the monopole.
The monopole density induced by the random spin-orbit interaction arises at the edge of the ferromagnetic film since
${\bm \nabla} \cdot {\bm N} \simeq \partial {N_x} / \partial x$ is finite there.
The induced monopole density at the two edges is
\begin{align}
\overline{\rho_{\rm m}}
=
{\mp} \frac{\beta_{\rm i}}{d} \overline{N},
\end{align}
where the sign is positive on one side of the edge and negative on the other side.
The monopoles then produces a magnetic field along the $x$ direction as
\begin{align}
\overline{{\bm B}_{\rm s}} = -{\bf e}_{x} \beta_{\rm i} \overline{N}.
\end{align}
This field creates the electric current in the $y$ direction via the conventional Amp\`ere's law.
The averaged electric current density generated by the spin damping monopole [Eq.~(\ref{jresult})] thus reduces to 
\begin{align}
\overline{\bm j}
=
-{\bf e}_y (\sigma_{\rm c} \alpha_{\rm R} E_{\rm R} +\frac{\beta_{\rm i}}{\mu d}) \overline{N}.  \label{javerageres}
\end{align}

We have thus found that when magnetization precession occurs in a metallic ferro-normal junction, a current of voltage arises perpendicular to the junction and a precession axis. 
As far as electric detection concerns, a monopole effect appears qualitatively the same as an inverse spin Hall effect \cite{Saitoh06}, which is a widely-used experimental method for spin current detection.
We will examine this important point later in \S\ref{sec:ishe}.

Let us look into the monopole signal quantitatively.
We define energy scales of Rashba and impurity spin-orbit interactions as 
($\kf$ is Fermi wavelength)
\begin{align}
\Delta_{\rm R} &\equiv \frac{\lambda_{\rm R}}{e}\kf^2 v_{\rm R} \nnr
\Delta_{\rm i} &\equiv \lambda_{\rm i}\kf^2 v_{\rm i},
\end{align}
where $v_{\rm R}\equiv \frac{e}{\kf}E_{\rm R}$ is a potential energy due to Rashba electric field, $\Ev_{\rm R}$.
Coefficients $\alpha_{\rm R}$ and $\beta_{\rm i}$ then read 
\begin{align}
\alpha_{\rm R} &= 12\pi^2 \lt(\frac{J}{\ef}\rt)^2 
\lt(\frac{\Delta_{\rm R}}{v_{\rm R}}\rt) \tau \nnr
\beta_{\rm i} &= \frac{16}{3}\mu e \kf 
\lt(\frac{J}{\ef}\rt)^2 
\lt(\frac{\Delta_{\rm i}}{v_{\rm i}}\rt)
\lt(\frac{\ef\tau}{\hbar}\rt)^2.
\end{align}
$\beta_i$ and $\mu e \kf$ have dimension of Tesla$\cdot$sec.
Here we used $\nu=\kf^3/\ef$, $n=\kf^3/(6\pi^2)$ and
$\sigma_{\rm c}=e^2 n\tau/m$.
 
When the spin damping arises from the magnetization precession with  the frequency $\Omega$ and the angle $\theta$ (Fig. \ref{fig;system}), the  magnitude of a monopole-induced current density, \Eqref{javerageres}, reads
\begin{align}
|\bar{j}| = 4e k_{\rm F}^2 \Omega \lt( \frac{J\tau}{\hbar} \rt)^2 \sin{\theta} \lt[\frac{\Delta_{\rm R}}{\varepsilon_{\rm F}} 
  +\frac{4}{3}\frac{1}{k_{\rm F}d} \frac{\Delta_{\rm i}}{v_{\rm i}}\rt].
\end{align}
We consider a disordered ferromagnets with 
$J / \varepsilon_{\rm F} \sim 0.1$, 
$\varepsilon_{\rm F}\tau / \hbar \sim 10$ and $k_{\rm F}^{-1} \sim 2$ \AA
($\ef=\frac{\hbar^2\kf^2}{2m}=1.51\times 10^{-19}$J$=0.95$eV and $\tau=7.0\times 10^{-15}$s).
Rashba interaction strength is chosen as 
$\Delta_{\rm R} / \varepsilon_{\rm F} \sim 0.1$, considering
an enhancement on the surfaces and the interfaces~\cite{Ast07}.
We assume for simplicity that 
$\frac{1}{\kf d}\frac{\Delta_{\rm i}}{v{\rm i}}$ is the same order of magnitude as $\frac{\Delta_{\rm R}}{\ef}$.
When $\theta = 30^\circ$ and $\Omega = 1$ GHz, the electric current density is thus 
\begin{align}
|\bar{j}|=1.6 \times 10^9 \;\;\rm{ A/m}^2,
\end{align}
 which is sufficiently large for experimental detection.
In addition to DC, there is AC component in Eq.~(\ref{jresult}) which would be accessible by a time-resolved measurement.

The current corresponds to an effective electric field, \Eqref{EBdef}, 
with a magnitude
\begin{align}
|E_{\rm S}| =\frac{1}{e}\alpha_{\rm R} \kf v_{\rm R} \Omega
  =12\pi^2\frac{\ef\kf}{e} \lt(\frac{\Delta_{\rm R}}{\ef}\rt)^2
\lt(\frac{J}{\ef}\rt)^2 \tau \Omega . \label{emf}
\end{align}
For the above values of parameters, the field is 
390 V/m.
The voltage induced by monopole when a sample width is 
100nm 
is thus
39$\mu$V.
The magnitude of effective magnetic field, \Eqref{EBdef}, is
\begin{align}
|B_{\rm S}| =\beta_{\rm i} \Omega
  =\frac{16}{3} \mu e \kf \Omega  \lt(\frac{\Delta_{\rm i}}{v_{\rm i}}\rt)^2
\lt(\frac{J\tau}{\hbar}\rt)^2 ,
\end{align}
which is estimated as
5.4m Tesla if we choose $\mu=10^5\times \muz$ ($\muz=4\pi\times 10^7$H/m is permeability in vacuum) as in the case of permeability of permalloy for the ordinary magnetic field.
We note that $\mu$ here is a permeability for effective magnetic field and can be different from the one for ordinary magnetic field.
For correct quantitative estimate, magnetic properties for the effective magnetic field needs to be investigated further.

The monopole current, \Eqref{jmdef}, is estimated to be
\begin{align}
|j_{\rm m}|\simeq \frac{|E_{\rm s}|}{d}\sim 4\times 10^{11}  \;\; {\rm V}/{{\rm m}^2}.
\end{align}
This value, however, cannot be converted into usual dimension of current density, A/m$^2$, since permeability of the effective field is not known.

We note that the electric current estimated here is an initial current that arises when the pumping of monopoles starts. 
When the monopole current is pumped steadily, the monopole accumulation grows at the edges of the system, inducing a diffusive current.
The steady monopole distribution is then determined by the balance of this backward diffusion and the pumped monopole current.

\section{Discussion}

In this paper, we have succeeded in proving existence of monopoles by deriving Maxwell's equation by a transport calculation in \S\ref{sec:hhmpweakcoupling} and \S\ref{sec:sdmp}. 
There remains, however, a few points to be clarified in the transport approach to spin damping monopole. 

First is a question whether spin damping monopole has a topological meaning or not.
In a case of hedgehog monopole, it was a topological object in three space dimensions, while topological nature seems to be lacking in the expression of spin damping monopole density (\Eqref{rhomdef}).
What is crucially different in spin damping monopole is that it needs dynamic spins ($\dot{\Sv}$). 
Thus, if there is any topological meaning, it should be discussed in both space and time.

Second point is a gauge field representation of spin damping monopole.
In a case of hedgehog monopole, it was obvious in the strong coupling limit.
Even in the weak coupling limit, it is straightforward to see from the effective fields (\Eqref{eq;fields}) that the gauge field describing hedgehog monopole is 
$\Av_{\rm S}^z=\frac{1}{2}  (1-\cos\theta)\partial_\mu \phi$ (defined in \Eqref{Asdef}).
A quantization condition for a monopole charge then arose naturally from the condition that a gauge field can be defined to cover the whole space (\Eqref{diracquantization2}).
In a case of spin damping monopole, an effective magnetic field in \Eqref{EBdef} cannot be written as a rotation of a local effective vector potential (because there is a monopole), but is represented by a nonlocal vector potential as ${\bm B}_{\rm s}=\nabla\times \Av^{\rm (sd)}+{\bm \Phi}$, where 
 (see \Eqref{AfromBstring}) 
\begin{align}
A_i^{\rm (sd)}(\rv)=-\beta_{\rm i}\sum_{jk}\epsilon_{ijk}\int_Ldr'_j 
{N}_k(\rv-\rv'), \label{Aforsdmp}
\end{align}
and ${\bm \Phi}$ represents a string singularity field. 
Based on this expression, the same argument as Dirac for a quantization condition \cite{Dirac31} is valid, and thus the charge of a spin damping monopole $g$ is quantized as in \Eqref{diracquantization}.
In order to understand fully the origin of spin damping monopole, it is important to study if the monopole is represented by patching two or more local gauge fields (see \S\ref{sec:quantizationhhmp}).  
This would be carried out by studying a clean ($\tau=\infty$) and strong sd coupling limit.

\subsection{Monopole interpretation of the inverse spin Hall effect
\label{sec:ishe}}

The electromotive force discussed above is a result of the Maxwell's equation for the monopole, which is an exact equation required by the gauge invariance.
If one detects the electromotive force given by Eq. (\ref{emf}), it becomes a direct evidence of spin damping monopole.
The electromotive force here acts on the spin of the electron, and thus can be detected electrically as has been demonstrated \cite{Yang09}. 
Remarkably, the observation of spintronics monopoles might have already been achieved.
In fact, the electric voltage due to the magnetization precession has been observed in a system of ferromagnet on a Pt film \cite{Saitoh06} and the direction of the voltage was in agreement with our prediction from the monopole. 
(The spin-orbit interaction expected in Pt is an intrinsic one due to periodic atoms, but the effect is expected to be the same as the one from the random potential. In addition, interface Rashba interaction might also be there.
Therefore the monopole scenario would apply to the system in Ref. \cite{Saitoh06}.)
In Ref. \cite{Saitoh06}, the mechanism for the voltage generation was argued to be the inverse spin Hall effect. 
According to the inverse spin Hall explanation, the magnetization precession generates spin current via the spin pumping effect, and the spin current, $\js$,  is converted into charge current by the spin-orbit interaction (the inverse of the spin Hall effect).
This explanation assumes that 
the conversion mechanism of  \cite{Hirsch99}
\begin{align}
j_i=\lambda \sum_{jk}\epsilon_{ijk} j_{{\rm s},j}^k,
\end{align}
where $\lambda$ is a constant representing the strength of the spin-orbit interaction  and $k$ is the index for the spin polarization of the spin current.
This formula is, however,  physically incorrect. 
In fact, the charge current is a conserved quantity, while the spin current is not because of the spin relaxation, and thus these two currents should not be simply proportional to each other.
To put in other words, spin current does not have a unique definition because of its relaxation, and thus is not physical.
A spin current explanation is therefore an approximate one which might be justified only at very short distance (less than the spin relaxation length).
Indeed, it has been theoretically demonstrated that that conversion formula is not satisfied in a case of slowly varying magnetization configuration \cite{Takeuchi10}.
In contrast to the spin current scenario,  
our monopole scenario is a result of Maxwell's equations, required by a symmetry of the electromagnetism.  
Since monopole is a conserved quantity satisfying $\dot{\rhom}+\nabla\cdot \jvm=0$, 
monopole scenario is based on a relation between physical quantities, the electric field and monopole.
Our scenario therefore explains the "inverse spin Hall effect" free from ambiguity, in contrast to the spin current one.

For experimental confirmation of the spin damping monopole, 
of crucial importance is the separation of the monopole signal from the inverse spin Hall signal driven by the spin current.
This is accomplished by applying an effective electric field perpendicular to the junction of Fig.~\ref{fig;system}.
The monopole contribution then leads to the transverse electric current as a result of the Hall effect for the monopole, while the contribution of the spin current is not affected.

Observation of the effective magnetic field emitted by monopoles, \Eqref{EBdef}, by use of muons or neutrons or electron holography
would be also a direct evidence of monopole generation.

\subsection{Remarks on spin current}

Our analysis has also proved that the spin current does not modify the fundamental law of electromagnetism, namely Maxwell's equations. 
Historically, there have been arguments that the spin current would create the electromotive force by a modified Amp\`ere's law based on the assumption that the spin current is equivalent to the flow of two monopole charges \cite{Sun04}.
Quantum mechanics tells us that this idea is too naive, since the spin is a point object and thus the separation between the magnetic charges are strictly zero, even if one dares to interpret spin by two magnetic charges. 
The monopole current associated with the spin current is therefore absolutely zero.
In fact, spin current is a quantity in an SU(2) spin space which couples to the electromagnetism only when properly projected, and since we know from the argument by Volovik \cite{Volovik87} that the projection in the adiabatic limit gives rise only to a hedgehog monopole current but not the spin current. 
Our study has proved that the correct projection of the magnetic systems with the spin-orbit interaction introduces a magnetic monopole in electromagnetism.

\section{Summary}

As we have discussed, magnetic monopoles are common objects in solids; they emerge in ferromagnetic metals from topological spin structures and from dynamic spin (magnetization) with damping. 
Although monopoles have been discussed so far based on a gauge theory, 
we have presented a novel method to access monopoles based on a transport calculation.

In monopoles in ferromagnets, of particular interest are  spin damping monopoles, generated in a simple system of a ferromagnet and heavy metal.
They open a novel path for connecting magnetism and electronics via an analog of Amp\`ere's law.
A novel concept of monopolotronics, control of monopoles, proposed in Ref. \cite{Takeuchi12}, is expected to be useful for realizing next generation spintronic devices.

\begin{acknowledgments}
The authors are grateful to N. Kitazawa for discussion.
This work was supported by 
a Grant-in-Aid for Scientific Research (B) (Grant No. 22340104) from Japan Society for the Promotion of Science 
and UK-Japanese Collaboration on
Current-Driven Domain Wall Dynamics from JST.
Two of the authors (A.T. and K.T.) are financially supported by the Japan Society for the Promotion of Science for Young Scientists.
\end{acknowledgments}

\appendix

\section{Electromagnetic fields in the relativistic representation }
\label{sec:relativistic}

In this section, we briefly summarize the electromagnetism theory in the relativistic representation using 4-vectors.
We follow the convention used in Ref. \cite{Ryder96}.
Contravariant vectors, such as $x^\mu=(t,x,y,z)$, are represented by the upper indices. 
Covariant vectors, represented by lower indices,  are defined by multiplying the metric tensor $g_{\mu\nu}$ as, for example, 
$x_\mu=g_{\mu\nu}x^\nu$, 
where 
\begin{align}
g_{\mu\nu}=\lt( 
  \begin{array}{cccc} 1 & 0 & 0 & 0 \\
                0 & -1 & 0 & 0 \\
                0 & 0 & -1 & 0\\
                0 & 0 & 0 & -1 \end{array} \rt).
\end{align}
Thus covariant vector of the 4-dimensional coordinate is thus $x_\mu=(t,-x,-y,-z)$.
The product of the covariant vector and contravariant vector is a Lorentz invariant scalar, e.g., 
$x_\mu x^\mu= t^2-\rv^2$.
(Note that we use relativistic notation only in this section. 
In other sections, upper and lower indices of the three-dimensional vectors means the same.)
Differential operators are
\begin{align}
 \partial_\mu\equiv \delpo{x^\mu}=\lt(\delpo{t},\delpo{x},\delpo{y},\delpo{z}\rt), 
 \partial^\mu\equiv \delpo{x_\mu}=\lt(\delpo{t},-\delpo{x},-\delpo{y},-\delpo{z}\rt).
\end{align}
We first describe the electromagnetism without monopole, namely, when there is.
The electromagnetic field tensor is defined by a U(1) gauge field $A^\mu$ as
$F^{\mu\nu}=\partial^\mu A^\nu-\partial^\nu A^\mu$.
Its components are (noting $\partial_i=-\partial^i$)
$F^{ij}= -\epsilon_{ijk}B^k=F_{ij}$, where $B^k=(\nabla\times \Av)^k$, 
and 
$F^{0i}=(\partial_t \Av+\nabla\phi)^i=-E^i=-F_{0i}$.
In the matrix representation, 
\begin{align}
F^{\mu\nu}=\partial^\mu A^\nu-\partial^\nu A^\mu
= 
  \lt(\begin{array}{cccc} 0 & -E_x & -E_y & -E_z \\
                E_x & 0 & -B_z & B_y \\
                E_y & B_z & 0 & -B_x \\
                E_z & -B_y & B_x & 0  \end{array} \rt).
\end{align}
The field strength tensor satisfies by definition the following identity
\begin{align}
\partial_\mu \tilde{F}^{\mu\nu}=0,\label{Bianchi}
\end{align}
where ($\epsilon^{0123}=1=-\epsilon_{0123}$ is the four-dimensional antisymmetric tensor)
\begin{align}
\tilde{F}^{\mu\nu}\equiv \frac{1}{2} \epsilon^{\mu\nu\rho\sigma}F_{\rho\sigma}. 
\end{align}
The 0-component of the identity (\ref{Bianchi}) is 
\begin{align}
\partial_i \tilde{F}^{i0}
=- \frac{1}{2} \epsilon^{ijk}\nabla_i F^{jk}=\nabla\cdot\Bv
=0,
\end{align}
and $i$-component reads
\begin{align}
\partial_0 \tilde{F}^{0i}+\partial_j \tilde{F}^{ji}
&= \frac{1}{2} \lt( \partial_0 \epsilon^{ijk} F_{jk}
-2\partial_j \epsilon^{ijk}F_{0k} \rt)  \nnr
&=-\delpo{t} \Bv - \nabla\times\Ev
=0.
\end{align}
Therefore, the condition of no monopole, $\nabla\cdot\Bv=0$, and the Faraday's law are trivial result of the U(1) gauge symmetry (definition of $F^{\mu\nu}=\partial^\mu A^\nu-\partial^\nu A^\mu$).

\section{Spin damping \label{sec:damping}}
Let us look into a role of spin damping represented by a vector $\Nv=\Sv\times \dot{\Sv}$.
A spin dynamics is induced by magnetic field.
An equation of motion for a spin is thus generally given by Landau-Lifshitz equation 
\begin{align}
\dot\Sv=\gamma \Bv_{\rm tot}\times \Sv,
\end{align}
where $\gamma$ is gyromagnetic ratio and $\Bv_{\rm tot}$ represents the total magnetic field acting on $\Sv$.
$\Bv_{\rm tot}$ thus includes an external magnetic field, an internal field due to exchange interaction with other localized spins, magnetic anisotropy field, and also the effect of coupling to other degrees of freedom such as conduction electrons and phonons.

We know that damping or dissipation arises in general from a coupling to an environment, namely, other degrees of freedom.
In a case of spins in metallic magnets, most strong source of dissipation is conduction electron.
Dissipation in this case is caused by spin-orbit interaction, which converts spin angular momentum into orbital angular momentum.
Therefore, damping effect is calculated by evaluating an effective magnetic field from conduction electrons, resulting in \cite{KTS06}
\begin{align}
\gamma \Bv_{\rm el} = -\alpha \dot\Sv,
\end{align}
where $\alpha$ is a constant proportional to the spin-orbit interaction.
The damping torque is thus represented by $\alpha\Sv\times \dot{\Sv}$, called the Gilbert damping term.
The equation of motion thus reads
\begin{align}
\dot\Sv=\gamma \Bv \times \Sv+ \alpha \Sv \times \dot\Sv , \label{LLGeq}
\end{align}
where $\Bv\equiv  \Bv_{\rm tot}-\Bv_{\rm env}$ is the field neglecting the effect of the environment, 
$\Bv_{\rm env}$.
Equation (\ref{LLGeq}) is called Landau-Lifshitz-Gilbert equation.

The effect of Gilbert damping torque is understood by looking into Fig. \ref{fig:damping}.
In fact, a vector $\Sv\times \dot{\Sv}\equiv \Nv$ tends to point the spin perpendicular to its precession direction, i.e., to the equilibrium direction along the field $\Bv$, and hence $\alpha\Nv$ represents the spin dissipated by the environment.


\bibliography{120305_icamd.bbl}

\end{document}